\documentclass[]{aa}
\usepackage{txfonts}
\usepackage{natbib}
\usepackage{amssymb}
\usepackage{amsmath}
\bibpunct{(}{)}{;}{a}{}{,}
\usepackage{epsf}
\usepackage{graphicx}
\usepackage{epsfig}
\usepackage{graphics}
\usepackage{subfigure}
\usepackage[english]{babel}
\usepackage{rotating}
\usepackage{color}
\usepackage[dvipsnames]{xcolor}

\newcommand{\mum}{$\mu$m}
\newcommand{\teff}{$T_{\rm{eff}}$}
\newcommand{\logg}{$\log g$}
\newcommand{\lL}{\ifmmode \log \frac{L}{L_{\sun}} \else $\log \frac{L}{L_{\sun}}$\fi}

\newcommand{\myr}{M$_{\sun}$ yr$^{-1}$}

\newcommand{\kms}{km~s$^{-1}$}

\newcommand{\lsun}{\ifmmode L_\sun \else L$_{\sun}$\fi}
\newcommand{\msun}{\ifmmode M_\sun \else M$_{\sun}$\fi}
\newcommand{\mstar}{\ifmmode M_\star \else $M_\star$\fi}
\newcommand{\hii}{H~{\sc ii}}
\newcommand{\lya}{\ifmmode {\rm Ly}\alpha \else Ly$\alpha$\fi}
\newcommand{\ha}{\ifmmode {\rm H}\alpha \else H$\alpha$\fi}
\newcommand{\magab}{mag$_{\rm AB}$}

\def\ergs{erg s$^{-1}$}
\def\ergscm{erg s$^{-1}$ cm$^{-2}$}

\begin{document}

\title{Spectral properties and detectability of supermassive stars in protoglobular clusters at high redshift \thanks{Result tables described in Appendix \ref{ap_photom} are available at the CDS via anonymous ftp to cdsarc.u-strasbg.fr}}
\author{F. Martins\inst{1}
\and D. Schaerer\inst{2,3}
\and L. Haemmerlé\inst{2}
\and C. Charbonnel\inst{2,3}
}
\institute{LUPM, Universit\'e de Montpellier, CNRS, Place Eug\`ene Bataillon, F-34095 Montpellier, France
\and
Department of Astronomy, University of Geneva, Chemin des Maillettes 51, 1290, Versoix, Switzerland
\and
IRAP, UMR 5277, CNRS and Université de Toulouse, 14, av. E. Belin, 31400, Toulouse, France
}

\offprints{F. Martins\\ \email{fabrice.martins@umontpellier.fr}}

\date{Received / Accepted }

\abstract
{Globular clusters (GCs) contain multiple stellar populations with peculiar chemical compositions. Pollution of the intracluster gas by an early population of fast-evolving stars is the most common scenario for explaining the observations. Stars with masses in excess of 1000 \msun\ have recently been suggested as potential polluters.}
{We investigate the spectral properties of proto-GCs that would host a supermassive star (SMS). Our main goal is to quantify how such a star would affect the integrated light of the cluster, and to study the detectability of such objects.}
{We computed nonlocal thermal equilibirum atmosphere models for SMS with various combinations of stellar parameters (luminosity, effective temperature, and mass) and metallicities appropriate for GCs, and we predict their emergent spectra. Using these spectra, we calculated the total emission of young proto-GCs with SMS as predicted in a previously reported scenario, and we computed synthetic photometry in UV, optical, and near-IR bands, in particular for the James Webb Space Telescope (JWST).} 
{At an effective temperature of 10000 K, the spectrum of SMSs shows a Balmer break in emission. This feature is due to strong nonlocal thermal equilibrium effects (implied by the high luminosity) and is not observed in "normal" stars. The hydrogen lines also show a peculiar behavior, with Balmer lines in emission while higher series lines are in absorption. At 7000 K, the Balmer break shows a strong absorption. At high effective temperatures, the Lyman break is found in emission. 
Cool and luminous SMSs are found to dominate the integrated spectrum of the cluster, except for the UV range. The predicted magnitudes of these proto-GCs are \magab $\sim$28-30 between 0.7 and 8 \mum\ and for redshifts $z \sim 4-10$, which is\ detectable with the JWST. The peculiar observational features of cool SMSs imply that they might in principle be detected in color-color diagrams that probe the spectral energy distribution below and above the Balmer break. 
 }
{Our results show that SMSs should be detectable in proto-GCs if they are luminous and relatively cool. They may be found through deep imaging with the JWST.}


\authorrunning{Martins et al.}
\titlerunning{Supermassive stars in globular clusters}

\maketitle


\section{Introduction}
\label{s_intro}
Globular clusters (GCs) are the oldest stellar systems known. Their ages reach up to the age of the Universe \citep[ e.g.,][]{2013ApJ...775..134V}. 
Initially thought to be simple compact clusters hosting an ensemble of stars with a uniform chemical composition formed in a single event, that is, a so-called single stellar population, GCs are now known to host multiple stellar populations (MSPs) that show very peculiar variations in chemical composition \citep[for reviews, see, e.g.,][]{1979ARA&A..17..309K,2012A&ARv..20...50G,2016EAS....80..177C,BL18}.

Triggered by these discoveries and the puzzles they pose, the interest in GCs has recently been revived, as testified by a flurry of new spectroscopic and photometric observations addressing their stellar content and the detailed chemical abundances of the MSPs  \citep[e.g.,][]{2009A&A...505..139C,2009A&A...505..117C,2010A&A...516A..55C,2009A&A...503..545L,2009ApJ...707L.190H,2015AJ....149..153M,2015MNRAS.450..815M,2018ApJ...859...81M,2019MNRAS.487.3815M,2015AJ....149...91P,2017MNRAS.464.3636M,2018MNRAS.481.5098M,2017A&A...601A.112P,2017MNRAS.466.1010S,2018ApJ...859...75M,2018MNRAS.473.2688M,2019A&A...622A.179B,latour19,2019A&A...622A.191M}.
A variety of scenarios have been proposed to explain the complex patterns emerging from the observations, in particular, the C-N, O-Na, and Mg-Al anticorrelations that result from hot hydrogen burning in short-lived polluter stars \citep{2007A&A...470..179P,pci17}. These scenarios invoke asymptotic giant branch stars \citep{2001ApJ...550L..65V,2012MNRAS.423.1521D}, 
fast-rotating massive stars \citep{2007A&A...464.1029D,2013A&A...552A.121K}, massive binaries \citep{2009A&A...507L...1D,2013MNRAS.436.2398B}, or recently, supermassive stars (SMS, with masses above a few $10^3$ M$_{\odot}$; \citealt{2014MNRAS.437L..21D,pci17,gieles18}), for example, to explain the observed variation in light elements.
A consistent model must also explain the formation of the MSPs, which requires a complex interplay between star formation in fairly extreme conditions, possibly nonstandard stellar evolution and initial mass function (IMF), and interstellar medium (ISM) physics and feedback \citep[e.g.,][]{2006A&A...458..135P,2008MNRAS.384.1231B,2008MNRAS.391..825D,2010A&A...516A..73D,2011MNRAS.413.2297S,2012A&A...546L...5K,2016A&A...587A..53K,2015ApJ...814L..14C,2016MNRAS.458.2122D,2016A&A...592A.111C,gieles18,2018MNRAS.476.2731V,2019MNRAS.486.2570B}. Despite important advances, it is fair to say that no agreement has yet been reached on the formation of GCs and their MSPs \citep[e.g.,][]{2015MNRAS.454.4197R,2015MNRAS.449.3333B,2016EAS....80..177C,BL18}. 

Significant progress has also been made in numerical simulations aimed at examining cluster formation and GCs in a cosmological context.
For example, \cite{Pfeffer2019The-evolution-o} and \cite{2019MNRAS.486.5838R} predicted the UV luminosities of young GCs, and examined the formation epoch of the GC populations as a function of their metallicity using the cosmological zoomed-in E-MOSAICS simulations. Similarly, \cite{2019arXiv190800984L} recovered systematic age differences of GCs with metallicity, as found by local observations. They also predicted that the most massive GCs form during major merger events. Despite these important advances and new insight that can be obtained from such simulations, none of them currently has  the mass and space resolution to properly describe the GCs, their complex structure, and the MSPs \citep{2018RSPSA.47470616F,2019ApJ...879L..18L,2019MNRAS.486.3134K}. 

Finally, observations have pushed the limits to very high redshift ($z \ga 6$), finding significant numbers of apparently very compact sources, some of which might be GCs in formation, which were observed in situ for the first time.
For example, using some of the deepest imaging data taken with Hubble in the Frontier Fields, \cite{Kawamata2015}, \cite{Bouwens2017Very-low-lumino}, \cite{2017MNRAS.467.4304V}, \cite{2019MNRAS.483.3618V}, and \cite{Kikuchihara2019Early-Low-Mass-} reported several such sources.
The sources are strongly magnified by gravitational lensing and remain basically unresolved in the HST images.
Based on the very small sizes ($\la 13-40$ pc) and absolute UV luminosities (down to $M_{\rm UV} \sim -15$), the authors estimated stellar masses of $\sim 10^6$ \msun\ or lower \citep{Bouwens2017Very-low-lumino,2019MNRAS.483.3618V}, which is in the range that is expected for GCs.
Although this is plausible, these compact rest-UV sources could also be the peaks of more extended objects whose full size remains undetectable with current instrumentation. Additional data and further evidence seem needed to definitely claim the observational detection of GCs in formation in the early Universe. Furthermore, clear criteria for identifying proto-GCs and distinguishing them from other ``normal" young clusters are required.

\citet{Renzini2017Finding-forming} and \citet{Pozzetti2019Search-instruct} have investigated the detectability of proto-GCs in the early Universe using evolutionary synthesis models. 
\citet{Pozzetti2019Search-instruct} accounted for `` normal" stars in their models, assumed a stellar mass of $2 \times 10^6$ \msun\ for a typical GC at formation, a metallicity of [Z/H]$=-1.35$ close to the mean of the GCs of the Milky Way, and predicted the detailed photometric properties of these objects. The authors showed in particular that these objects should be detectable at magnitudes $m_{\rm AB} \sim 30$ with the James Webb Space Telescope (JWST).
Although these calculations, based on simple stellar populations (hereafter SSPs), may be good guides for predicting the observable properties of young GCs, we know that globulars are more complex systems that host MSPs. It is  therefore important to examine the implications this may have on the integrated observational properties, in particular, in the context of searches for GCs in the early Universe, which now seem feasible.

We here explore one specific scenario for the formation of GCs and their MSPs: the SMS + proto-GC model of \cite{gieles18}. This scenario envisages the concurrent formation of a massive cluster and an SMS. This central object grows through runaway collisions and rapidly produces and ejects H-burning ashes that are incorporated into accreting protostars, producing second-population (2P) stars with peculiar abundance ratios, as observed in GCs. The scenario overcomes the mass-budget problem that is raised by the high ratio of second- versus first-population (1P) stars in all GCs \citep{2004ApJ...611..871D,2006A&A...458..135P,2011MNRAS.413.2297S,2012ApJ...758...21C,2015MNRAS.452..924K}. It also explains the observed increase in the fraction of 2P stars and of the He enrichment with the present-day mass of GCs \citep{2010A&A...516A..55C,2017A&A...601A.112P,2017MNRAS.464.3636M,2018MNRAS.481.5098M}. 

Observationally, an SMS, that is,\ a central stellar source of very high mass (M $\sim 10^3$ to $10^{4.5}$ \msun) and very high luminosity ($\sim 10^{7.5}$ to $10^9$ \lsun), may significantly alter the integrated properties of these GCs, if they exist.  To examine this question and explore the observational implications of the SMS scenario for the formation of GCs, we first predict the spectrophotometric properties of single SMSs using a dedicated nonlocal thermal equilibrium (non-LTE) stellar atmosphere model. Then we investigate {\em 1)} whether proto-GCs hosting a SMS would be detectable at high redshift, and {\em 2)} whether and how proto-GCs hosting SMS can be distinguished from ``normal" clusters (that lack SMS).

In other contexts, the spectral properties and detectability of metal-free (primordial, so-called Pop III) SMS have been studied by several groups. For example, \cite{2012ApJ...750...66J} have examined H and He recombination line emission from hot supermassive Pop III stars using some semianalytical models. Alternatively, \citet{Surace2019On-the-detectio,Surace2018On-the-Detectio} have investigated the spectrophotometric properties of primordial single SMS. They conclude that such objects, which are potential seeds of supermassive black holes
\citep[e.g.,][]{rees1978,volonteri2010,woods2019},
would be best detected if they were relatively cool. 
\cite{gieles18} have argued that the spectra of SMS in proto-GCs could resemble those of luminous blue variable (LBV) stars. 

We present the first detailed non-LTE atmosphere model calculations for SMS stars with non-zero metallicity, appropriate for GCs, which allows addressing the questions raised above.
Armed with spectral predictions for SMS, we can then quantitatively predict the total (integrated) spectra expected for proto-GCs that host SMS. 
The predictions made in this paper are expected to provide new criteria for searching and identifying young GCs with SMS in formation in the early Universe. If sources that show identifiable features of SMS can be found, this would also represent an important  test or confirmation of the SMS model for GC formation \citep{gieles18}.

The paper is structured as follows.  In Sect.\ \ref{s_sms} we summarize the expected properties of the SMS and describe the choice of parameters we explored for the non-LTE atmosphere models. The predicted spectra of SMS are presented and discussed in Sect.\ \ref{s_sms_prop}. In Sect.\ \ref{s_starclu} we derive the integrated spectra of young proto-GCs that contain both an SMS and a surrounding stellar population, we predict their photometric properties, and show how their spectral energy distributions (SEDs) can be distinguished from those of normal clusters in certain cases. Our results are discussed and compared to those from other studies in Sect.\ \ref{s_mass}. We finally summarize our main conclusions in Sect.~ \ref{s_conc}.

\section{Choice of parameters for SMS and clusters}
\label{s_sms}

\subsection{Formation scenarios and properties of SMS}

To compute the spectral properties of SMS, we need some estimates of their luminosity, effective temperature, and surface gravity. 
\citet{gieles18} used extrapolations from observed properties of massive stars to depict the mass-radius relation and the mass-loss rate of SMS and to predict their mass growth and their global properties as a function of the cluster mass. 
No stellar evolution models exist for SMS that formed through runaway collisions (in the metallicity range of GCs, the only existing SMS models start on the zero-age main sequence (ZAMS) and evolve at constant mass; \citealt{2014MNRAS.437L..21D}), and most studies are devoted to the case of Pop III SMS as seeds for supermassive black holes in galactic centers.  

Two categories of Pop III SMS models exist in the literature: monolithic models, where all the mass is assumed to have assembled at once
\citep{hoyle1963a,hoyle1963b,fowler1966,fuller1986,baumgarte1999a,baumgarte1999b,butler2018,dennison2019},
and accreting models, where the star grows supermassive through continuous accretion of gas at the surface
\citep{begelman2010,hosokawa2013,sakurai2015,sakurai2016a,umeda2016,woods2017,haemmerle2018a,haemmerle2018b,haemmerle2019a}.
Without external perturbative effects, monolithic models are assumed to be thermally relaxed, which implies surface properties corresponding to ZAMS models, that is, $T_{\rm eff}\sim10^5$ K \citep{ds02}. Stars that are near the Eddington limit and thermally relaxed must be mostly convective \citep{begelman2010}, which can be anticipated by extrapolation from the size of the convective core of conventional massive stars ($\lesssim100$ M$_\odot$), and is confirmed by numerical models (Woods, Heger \& Haemmerl\'e, in prep.).
 On the other hand, models accounting for continuous accretion indicate that above a threshold in the accretion rate ($\sim0.01$ M$_\odot$ yr$^{-1}$), the star cannot relax thermally because the thermal processes become inefficient on the short accretion timescale \citep{hosokawa2013,haemmerle2018a}.
As a consequence, its surface properties do not converge to ZAMS properties, but to those of red supergiants
on the Hayashi limit, with $T_{\rm eff}\sim5000-6000$ K (Fig.~\ref{fig_ML}).
Moreover, due to the inefficiency of entropy losses, accretion builds up an entropy profile that increases outward. Thus, the gas is stable with respect to convection, and the star remains radiative for 90\% of its mass.

These two categories of models are obvious simplifications, in particular in the context of GCs,
where stellar collisions are expected to be frequent and runaway collisions can be triggered \citep{gieles18}.
In dense environments, stellar mergers can contribute significantly to the mass-growth of SMSs
\citep{portegies1999,portegies2004a,freitag2006b,lupi2014,katz2015,boekholt2018,tagawa19}.
The impact of stellar mergers on the structure of SMSs is not known.
Such highly dynamical events might critically modify all the properties of SMSs,
in particular by redistributing entropy in the stellar interior.
If high-entropy gas can penetrate into deep regions of SMSs during mergers,
the resulting structure might differ significantly from that of accreting models and might
enhance convection and/or allow the surface to relax thermally.
Thus, while accretion and collisions must both play a significant role in the formation of SMSs at the center of GCs,
it is not clear if the properties of these objects correspond to the pure accretion models described above.
In the absence of suitable and reliable theoretical models,
observations could provide useful constraints on the respective roles of accretion and mergers.

\subsection{Choice of stellar parameters}
\label{s_sms_meth}

\begin{figure*}[]
\centering
\includegraphics[width=0.49\textwidth]{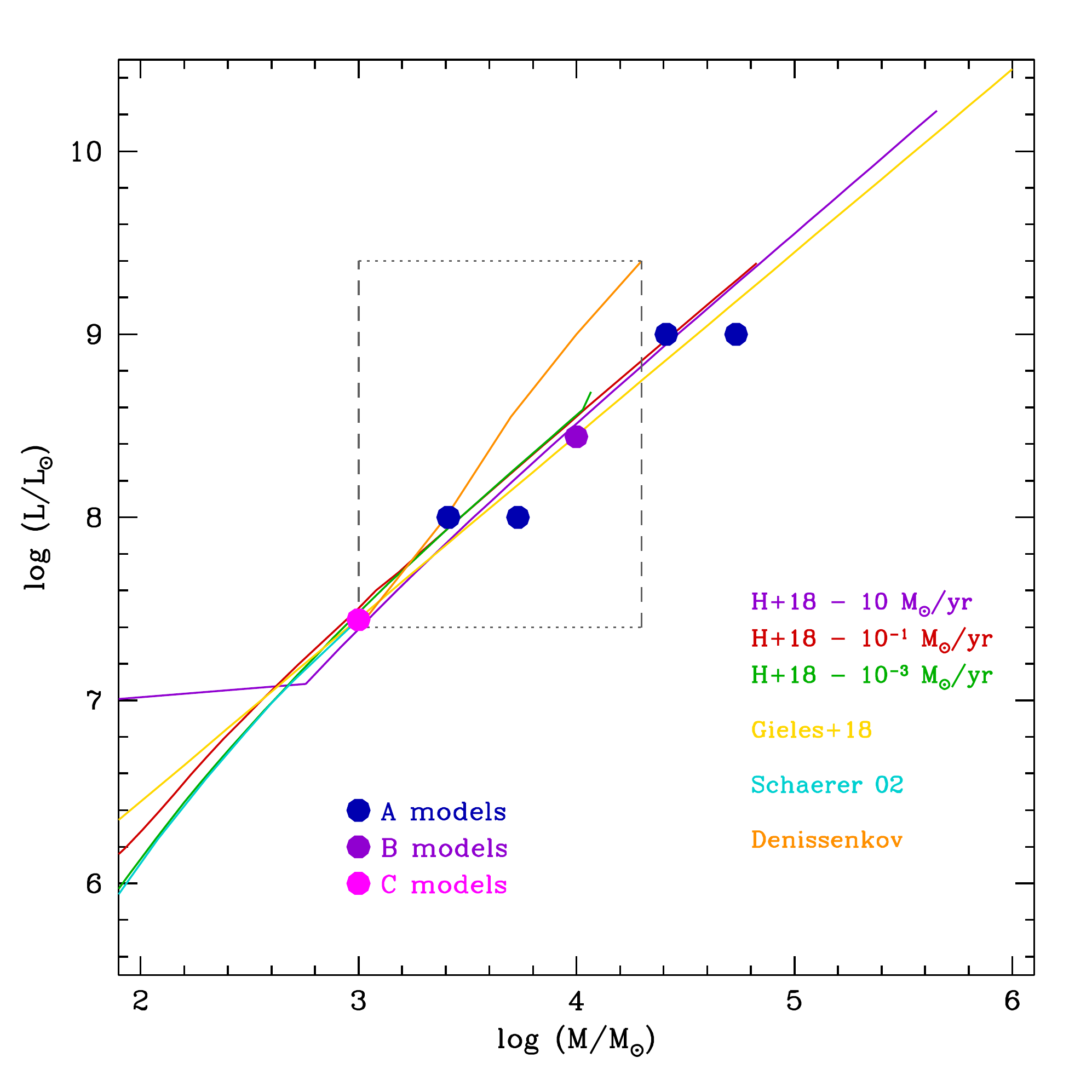}
\includegraphics[width=0.49\textwidth]{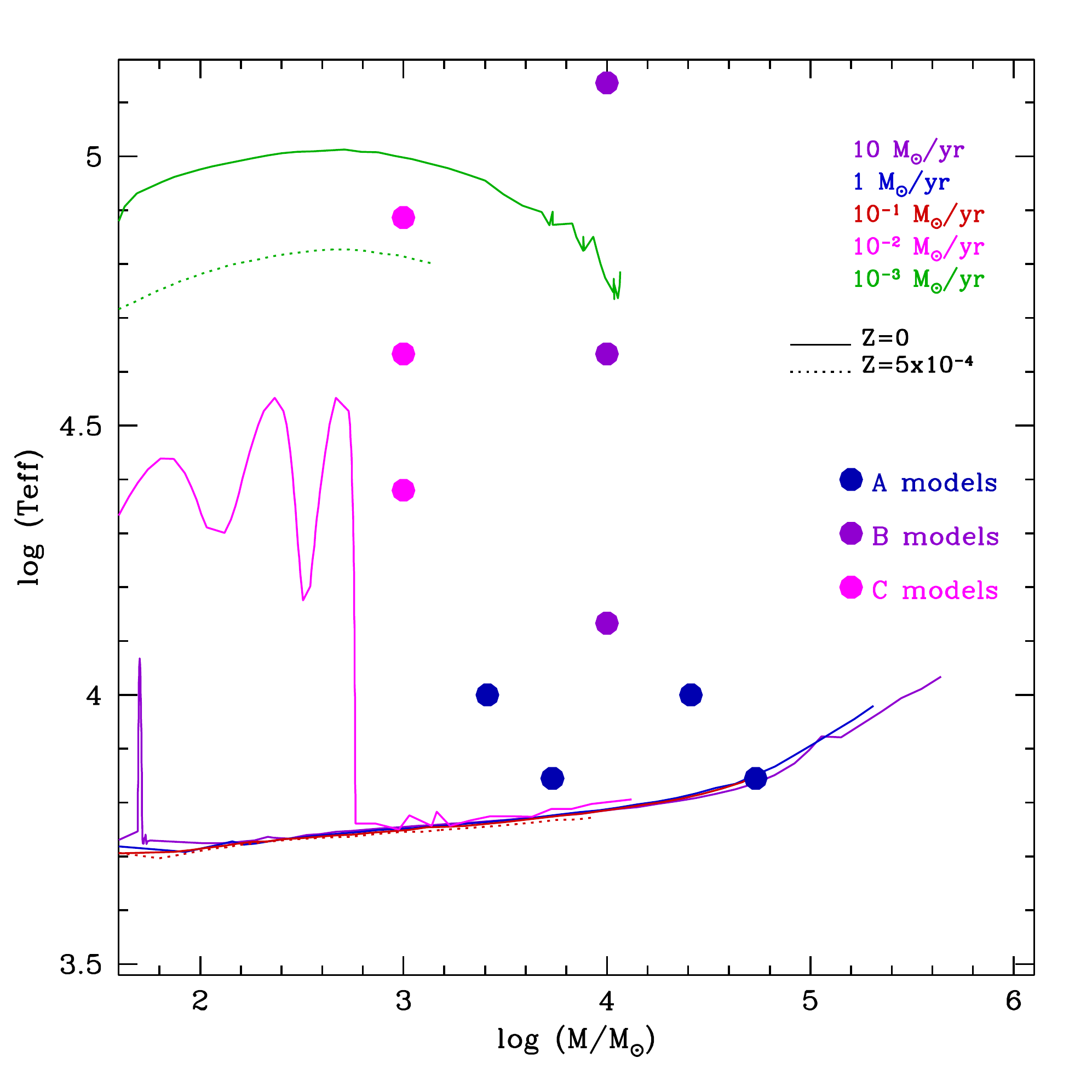}
\caption{\textit{Left}: Mass-luminosity relation (in solar units) adopted by \citet{gieles18}  compared to the predictions of the Pop III models of \citet{ds02} and of \citet[][- H+18]{lionel18} for three different accretion rates  and of the Pop II ZAMS models of Denissenkov et al.\ (priv. comm. see \citealt{pci17}). 
The vertical dashed lines show the mass limits that are allowed for nucleosynthesis; see text. The horizontal dotted lines show the minimum and maximum luminosities in this mass interval, according to all M-L relations.  \textit{Right}: Mass-effective temperature relations of \citet{lionel18} for various accretion rates and two metallicities. In both panels the filled circles show the positions of our models.}
\label{fig_ML}
\end{figure*}

A key constraint for our models is the mass of the SMS. For nucleosynthesis to be able to produce the correct yields of the various elements observed in GCs, \citet{pci17} have shown that the most relevant mass range of the SMS is 10$^3$ to $2 \times 10^4$ \msun\ (see their Fig.~4). From this mass interval, we can estimate the luminosity range of the SMS using mass-luminosity relations found in the literature. This is shown in Fig.\ \ref{fig_ML}, where we see that \lL\ ranges between 7.4 and 9.4, depending on the models. 

To compute synthetic spectra, we need to set the effective temperature, luminosity, and surface gravity (or equivalently, the mass). We used two categories of stellar parameters (models A versus B and C in Table\ \ref{tab_param}).

In the first category of models (A1 to A4), we assumed two values of \teff: 7000 and 10000~K. They correspond to relatively cool stars that seem to be favored by 1) formation by accretion, at rates high enough to ensure a rapid build-up of the star, and 2) runaway collisions in dense clusters ("cool" stars have larger radii that favor collisions). The temperatures we adopted are somewhat higher than advocated by \citet{lionel18}, for instance (they predict values of 5000-6000 K, see Fig.\ \ref{fig_ML}). However, at these temperatures, molecules appear in a stellar atmosphere, and these species are not handled by \emph{CMFGEN}. Alternatively, no atmosphere code has been designed so far for cool stars that includes both sphericity and a full non-LTE treatment. These two ingredients are mandatory for calculating SMS spectra given their large extensions and high luminosities.  
For these models, we also adopted two luminosity values (10$^8$ and $10^9$ L$_{\odot}$). These luminosities are representative of the luminosity range described above, see Fig.\ \ref{fig_ML}. We adopted a surface gravity \logg\ = 0.8 (10000~K models) and 0.5 (7000 ~K models), corresponding to masses of $\sim$2500 to $\sim$ 54000 \msun. For the A2 and A4 models (those with \lL\ = 9.0), a mass lower by a factor 2 to 3 would be required to match the constraint imposed by nucleosynthesis arguments. However, for lower masses the surface gravity is also lower, and radiation pressure overcomes gravity in the atmosphere. Consequently, no solution of the hydrodynamic structure is achievable. In other words, the Eddington limit is exceeded. Our models therefore work with the lowest possible gravity. In spite of this "mass discrepancy", the shape of the SED of models A2 and A4 is expected to be representative of the theoretically less massive models (see Sect.\ \ref{s_coolsms}). In particular, as we show below, the peculiar Balmer break morphology is accounted for.  

In the second category of models (B1-B3 and C1-C3 in Table\ \ref{tab_param}), we adopted the stellar parameters according to the prescriptions of \citet{gieles18}:
$R = 30\ R_{\odot}\ (\frac{M}{100 M_{\odot}})^{\delta}$ and 
$L = 2.8 \times\ 10^6 L_{\odot} (\frac{M}{100 M_{\odot}})$, with M, R, and L the SMS mass, radius, and luminosity. \teff\ follows directly from R and L.
We recall that these prescriptions are based on extrapolations of the properties of massive stars (M$\sim$100 \msun) in our Galaxy and the Magellanic Clouds. The parameter $\delta$ is arbitrary and was assumed to extend between 0 and 1 by \citet{gieles18}.
The mass of the SMS was set to either 10$^3$ or 10$^4$ \msun\ based on the nucleosynthesis arguments given above. For our computations, we chose $\delta$=0, 0.5, and 1 for M=10$^3$ and 10$^4$ \msun. 

The stellar parameters of our models are summarized in Table \ref{tab_param} and the resulting SEDs are shown in Fig.\ \ref{fig_sed}. For models of the A serie, we calculated an effective $\delta$ parameter by inverting equation 5 of \citet{gieles18} in Table 1, that is, $\delta = \frac{log(R/30)}{M/100,}$ where $R$ and $M$ are the radius and mass of the SMS in solar units. 

The predictions of SMS formation by runaway collisions  performed by \citet{gieles18} indicate that $\delta$=1.0 leads to the highest ratio of SMS and cluster mass. In their Fig.\ 3, a $10^4$~\msun\ SMS is formed within a few million years in a cluster with $10^6$ stars (and a mass of about 5 $10^5$ \msun\ after 3 Myr). This is understood because of the larger radius of stars with high $\delta$. The cross section of collisions with normal stars is thus higher. $\delta$ = 1.0 is thus the most favorable case for the detection of an SMS in a proto-GC (see Sect.\ \ref{s_starclu}). We also note that our models A1-A4 correspond to high $\delta$ values and are therefore included in $10^6$ stars cluster simulations.

\begin{table}
\begin{center}
  \caption{Stellar parameters adopted for our  models. For series A models, \teff\ is adopted. \teff\ of models for the B and C series is derived from the prescriptions of \citet{gieles18}. See text for details.} \label{tab_param}
\begin{tabular}{llccccc}
\hline
ID & \teff\ & \lL\    & \logg\  &  M            & R$^1$    & $\delta$\\    
  K  &         &         & [M$_{\odot}$] & [R$_{\odot}$]  \\
\hline
A1 & 7000   &  8.0    & 0.5     & 5395  &  6723 & 1.35$^2$ \\
A2 & 7000   &  9.0    & 0.5     & 53956 &  21506 & 1.05$^2$ \\
A3 & 10000  &  8.0       & 0.8     & 2585  &  3340 & 1.45$^2$\\
A4 & 10000  &  9.0       & 0.8     & 25848 &  10580 & 1.06$^2$ \\
\hline
B1 & 137000 &  8.4    & 5.5     & 10000 &  30   &  0 \\
B2 & 43000  &  8.4    & 3.5     & 10000 &  301  & 0.5 \\
B3 & 13600  &  8.4    & 1.5     & 10000 &  3006 & 1 \\
C1 & 77000  &  7.4    & 4.5     & 1000  &  30 & 0 \\
C2 & 43000  &  7.4    & 3.5     & 1000  &  95 & 0.5 \\
C3 & 24000  &  7.4    & 2.5     & 1000  &  305 & 1 \\
\hline                                                             
\end{tabular}                                                   
\tablefoot{1 R is the stellar radius at an optical depth equal to 2/3. 2 $\delta$ is calculated from mass and radius (see text).}
\end{center}                                                     
\end{table}

\subsection{Method for spectral synthesis}
\label{s_spec_meth}

We used the code \emph{CMFGEN} to compute atmosphere models and synthetic spectra. A full description of the code is presented in \citet{hm98}. Briefly, \emph{CMFGEN} solves the radiative transfer and statistical equilibrium equations in non-LTE conditions. Spherical geometry is adopted to take the extension due to stellar winds into account. The density structure is given as an input from a velocity structure and the mass conservation equation. In practice, the velocity structure is the combination of a pseudo-photospheric structure and the so-called $\beta$-velocity law. The former is computed under the assumptions of hydrostatic equilibrium and LTE. Tabulated Rosseland opacities are used as input. This inner structure is connected to the velocity law $v=v_{\infty} (1-r/R)^{\beta}$, where we assumed $\beta$=1.0, $v_{\infty}$ is the velocity at the top of the atmosphere, and $R$ is the stellar radius. \emph{CMFGEN} takes line-blanketing into account with a super-level approach. In our computations, we included H, He, C, N, O,  Si, S, and Fe. A 1/100 solar metallicity was adopted. The mass-loss rate was set to $10^{-5}$ \myr\ and the terminal velocity (i.e., velocity of the wind at the top of the atmosphere) to 500 \kms. These values are typical of winds of very massive stars \citep{vink18}. After the atmosphere model converged, a formal solution of the radiative transfer equation was performed to yield the synthetic spectrum between 10 \AA\ and 20 $\mu$m.

\section{Spectra of supermassive stars}
\label{s_sms_prop}

\begin{figure*}[t]
\centering
\includegraphics[width=0.42\textwidth]{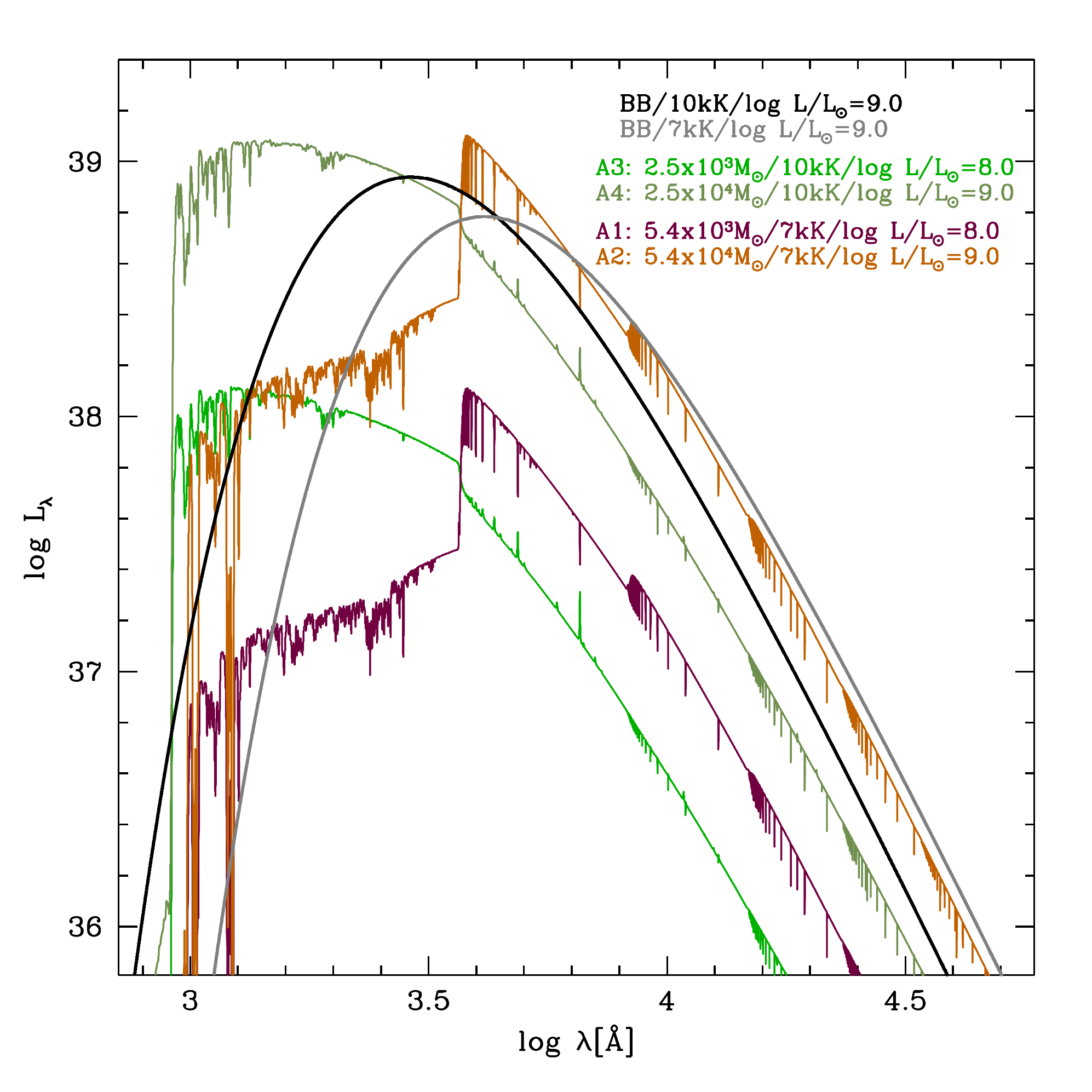}
\includegraphics[width=0.42\textwidth]{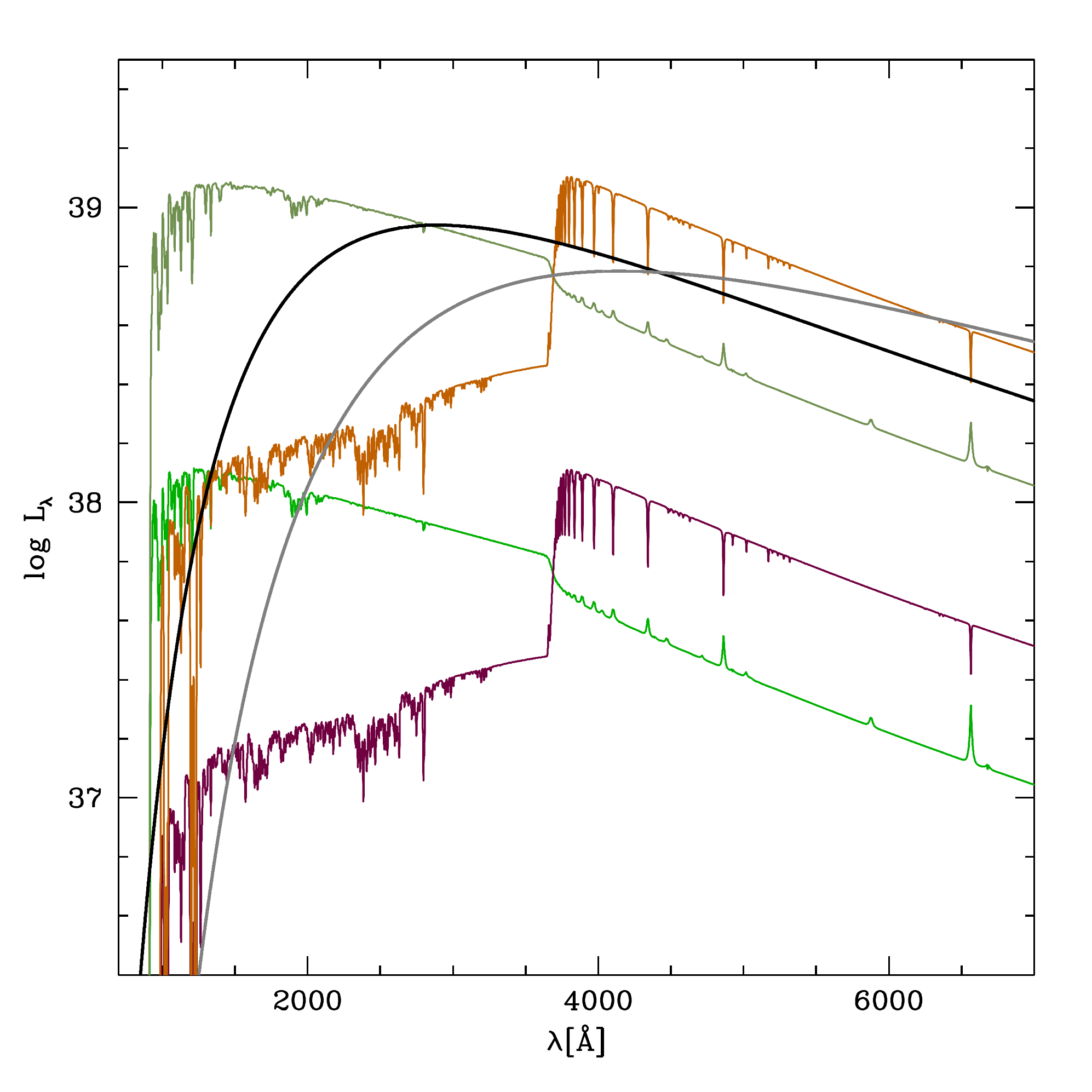}\\
\includegraphics[width=0.42\textwidth]{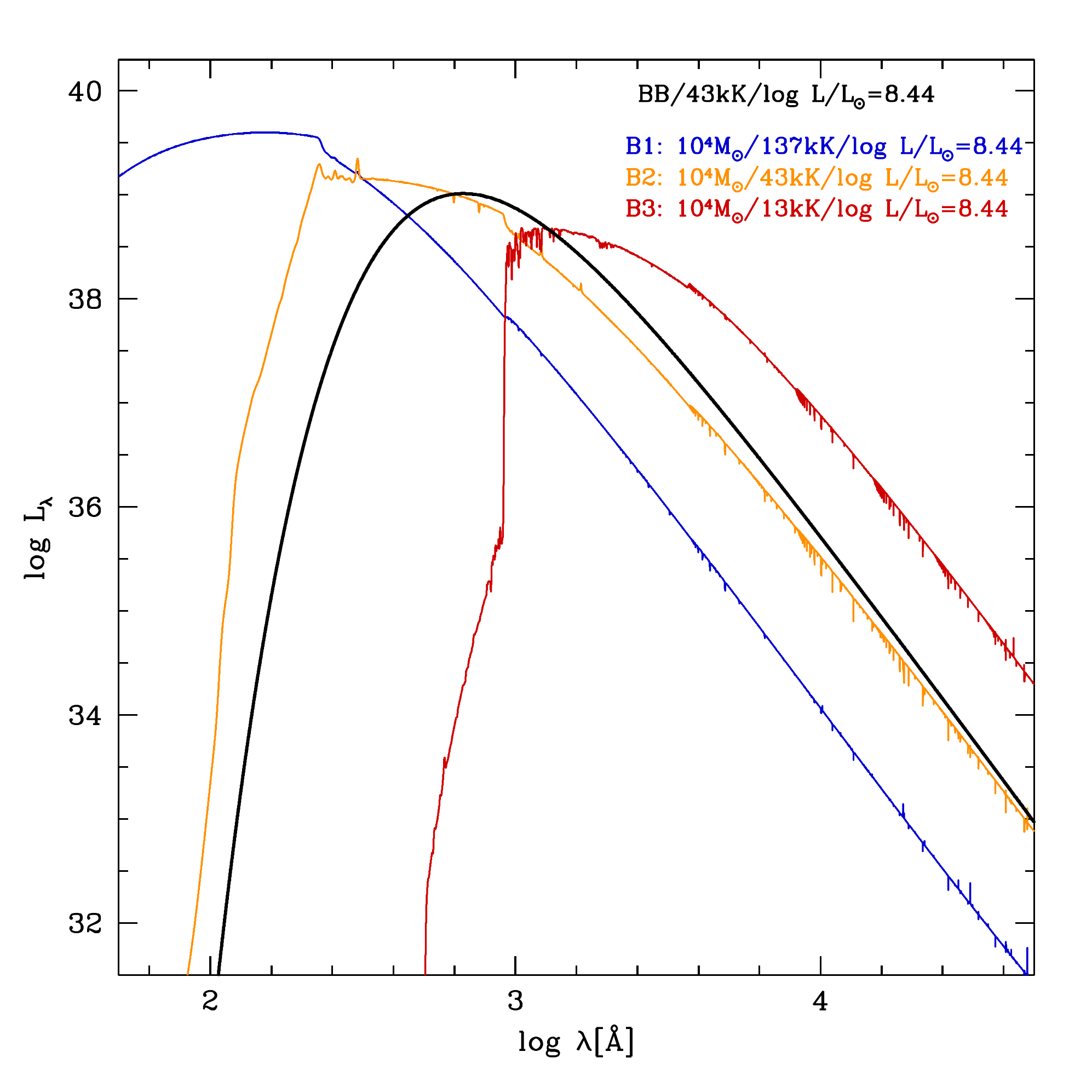}
\includegraphics[width=0.42\textwidth]{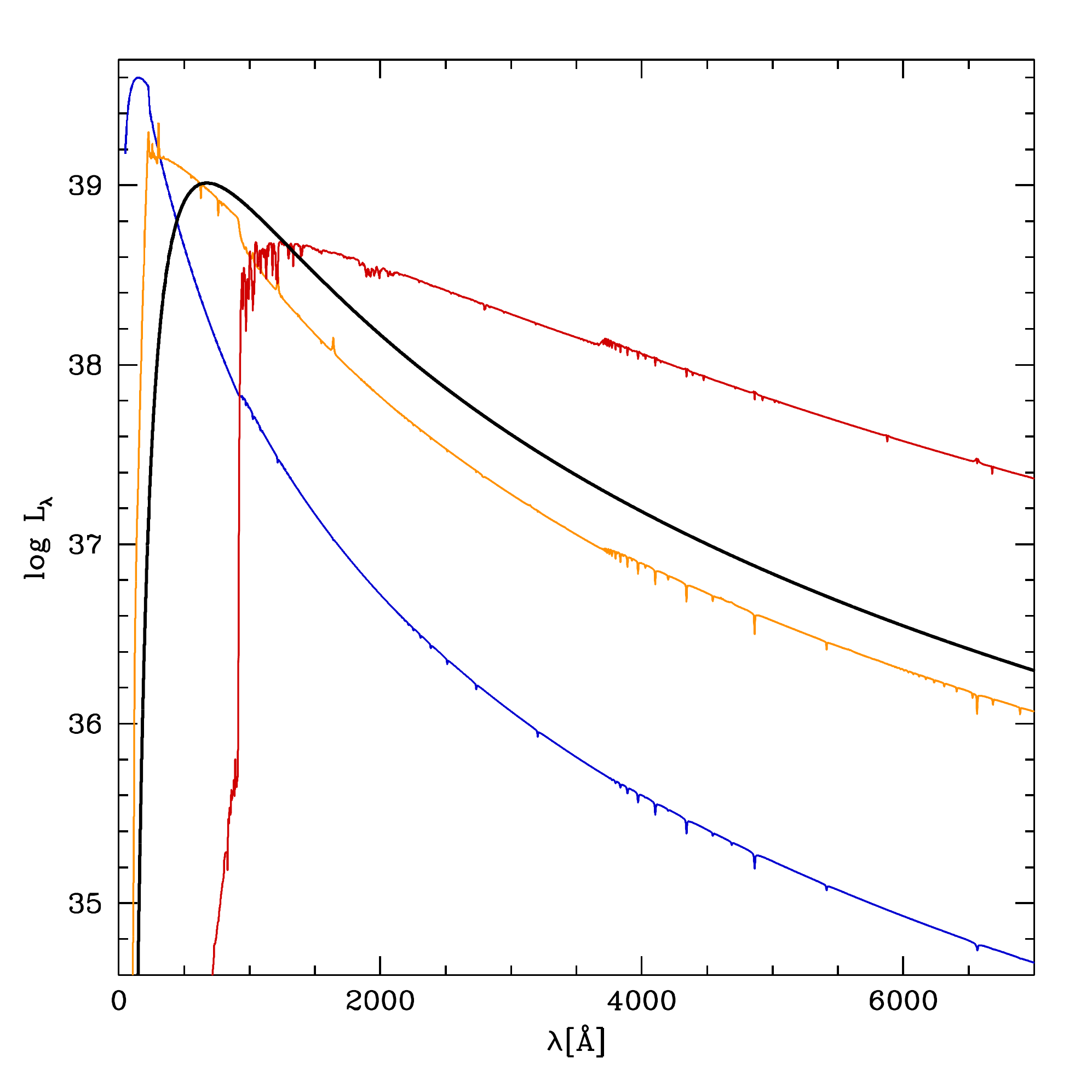}\\
\includegraphics[width=0.42\textwidth]{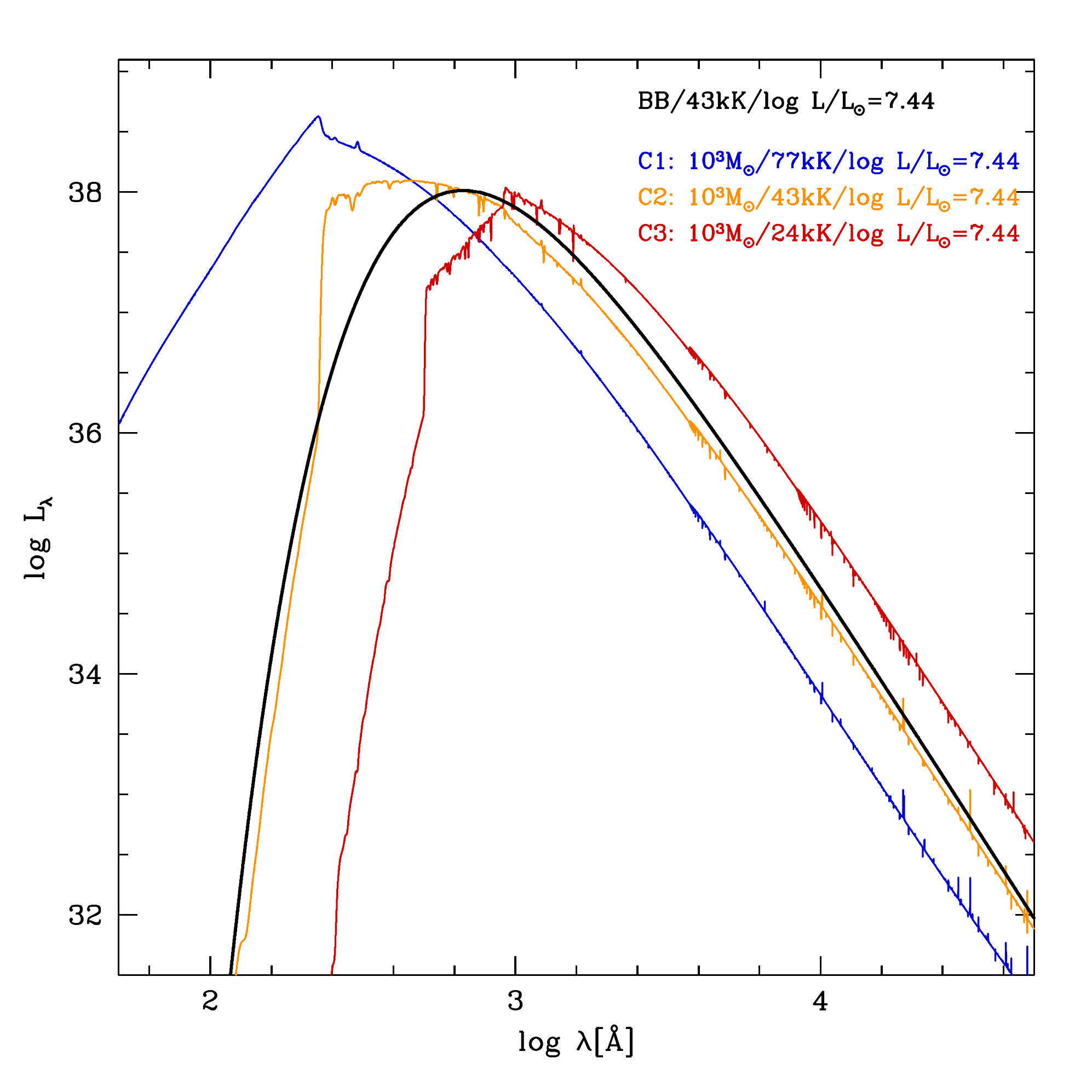}
\includegraphics[width=0.42\textwidth]{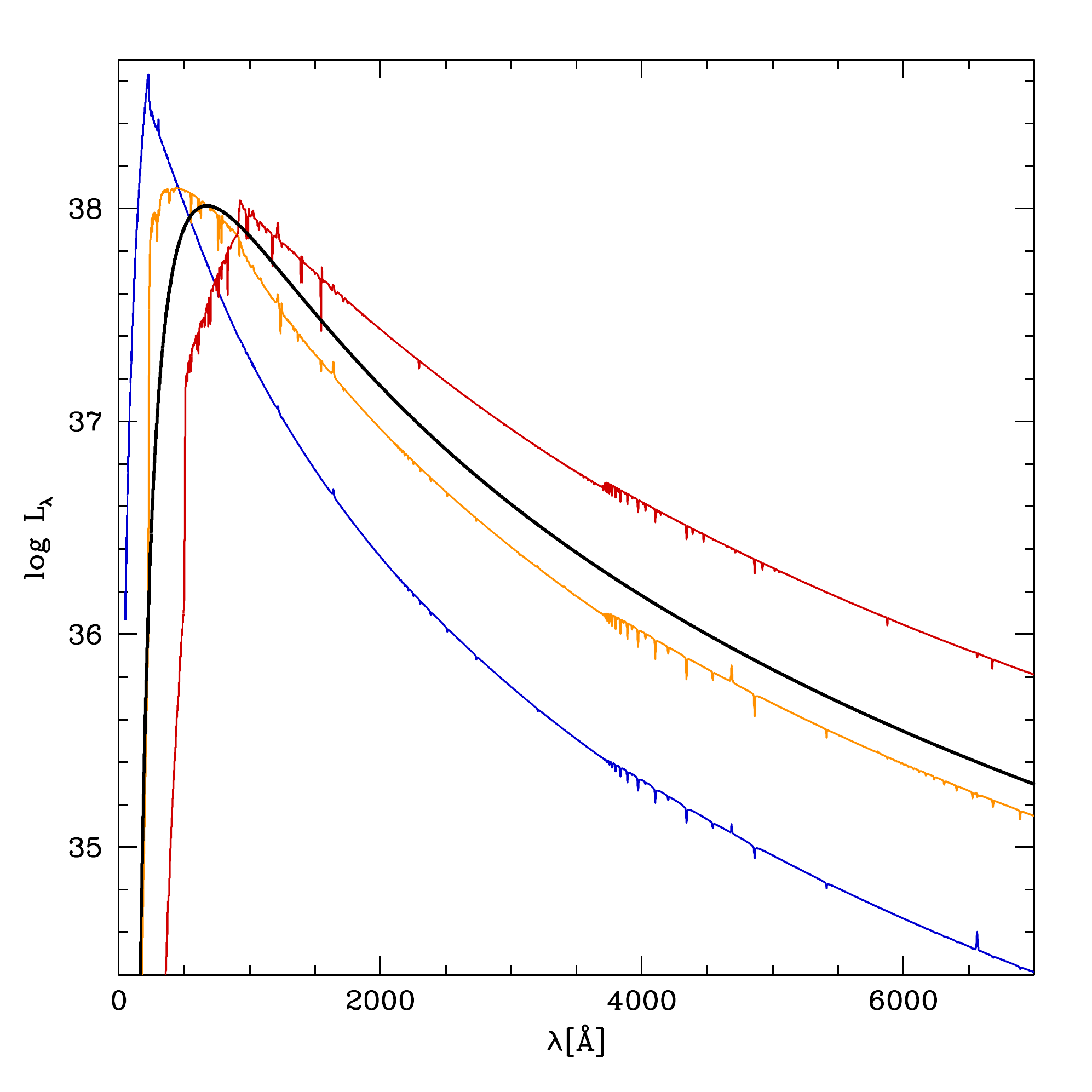}\\
\caption{Spectral energy distribution (in units of erg/s/\AA) for the models whose stellar parameters are summarized in Table \ref{tab_param}. The right panels are a zoom on the UV-optical part of the spectrum. The stellar parameters of each model are given in the left panel. The black and gray lines are blackbody spectra.}
\label{fig_sed}
\end{figure*}

An overview of all the SMS predictions is shown in Fig.\ \ref{fig_sed}, where
the spectra are compared to blackbody spectra with corresponding \teff. In the high-temperature range, a blackbody overestimates the stellar emission in most of the wavelength range, except near the emission peak in the far-UV. Qualitatively, the same trend is observed at 10000 K, where the blackbody emission is lower than the model emission only below the Balmer break. For the 7000 K models, the \emph{CMFGEN} model dominates shortward of the Balmer jump, while the blackbody flux is higher in the near-infrared and at longer wavelengths.
As expected, blackbodies are clearly not a good approximation for the predicted spectra of SMS. As we show below, some spectral peculiarities due to the high luminosity prevent a simple scaling of SEDs from normal stars. 
We now discuss the different SMS spectra, deviations from non-LTE, and the main physical effects affecting them.

\clearpage

\subsection{Cool SMS} 
\label{s_coolsms}

\begin{figure}[t]
\centering
\includegraphics[width=0.49\textwidth]{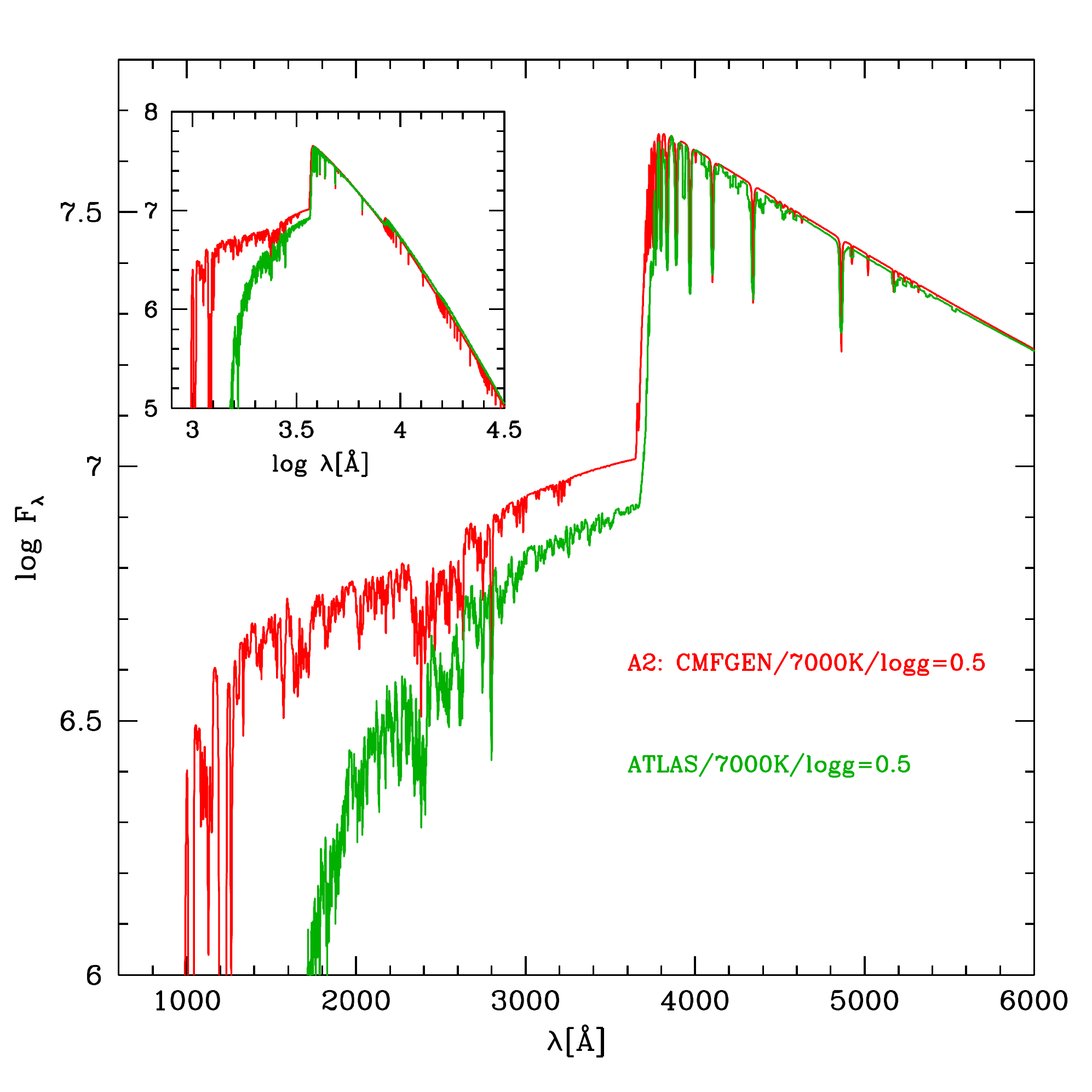}
\caption{Comparison between the SED of our model A2 and an \emph{ATLAS} model with \teff\ = 7000~K and \logg\ = 0.5. The fluxes are given in erg/s/\AA/cm$^2$.}
\label{fig_compatlas}
\end{figure}

The coldest models we considered for the cool SMS with \teff $\le 10000$ K (A1 and A2) are the only models that show a very strong Balmer break (see upper right of Fig.\ \ref{fig_sed}), which is\ the classical behavior of an A-type spectrum. 
As shown in Fig.\ \ref{fig_compatlas}, the predictions from our non-LTE models A1 and A2 are rather similar to the spectrum predicted from LTE models that were computed with the atmosphere code \emph{ATLAS} \citep{ck03}.
In the optical part between the Balmer and Paschen breaks, both types of models predict almost the same flux level. Below the Balmer break (3646 \AA), flux in the \emph{CMFGEN} model is at least 1.25 times higher and more than three times higher below 2000 \AA. Above the Paschen break (8204 \AA), the flux in the \emph{ATLAS} model is higher by about 10\% than that of the \emph{CMFGEN} model. 

\begin{figure}[t]
\centering
\includegraphics[width=0.49\textwidth]{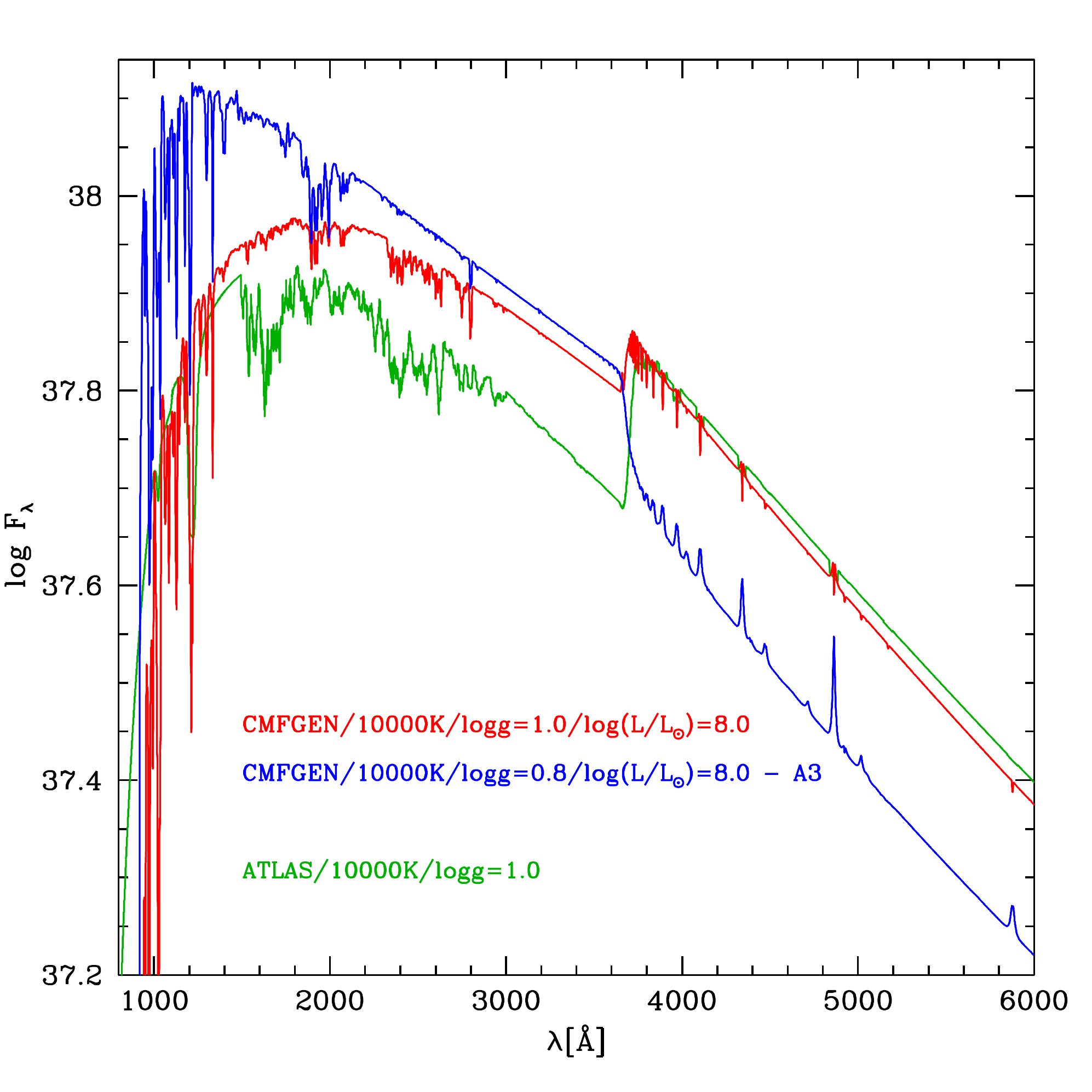}
\caption{Spectra of two models with \teff\ = 10000 K, \lL\ = 8.0 and \logg\ = 0.8 (blue) and 1.0 (red). The green spectrum is an \emph{ATLAS} model with \teff\ = 10000 K and \logg\ = 1.0.}
\label{fig_spec_Balmer}
\end{figure}

Fig.\ \ref{fig_spec_Balmer} shows the UV-optical part of the A3 model together with a comparison model at \logg\ = 1.0. In the former the Balmer jump is in emission, and the latter shows a more classical absorption shortward of 3646 \AA, but less pronounced than in models A1 and A2 discussed above. Fig.\ \ref{fig_spec_Balmer} also shows an LTE model computed with the \emph{ATLAS} code. Compared to the corresponding \emph{CMFGEN} model, the strength of the Balmer break is severely modified; it is much stronger in the LTE model. In the case of the SMS spectra (models A3 and A4), we even predict a Balmer break in emission, which cannot be reproduced by LTE models (see below). Adopting this LTE model for SMS is thus a very crude approximation that would underestimate the flux shortward of the Balmer break.

In Appendix \ref{ap_Balmer_logg} we show the effect of \logg\ on the shape of the SED for additional models with \teff\ = 10000 K. The Balmer break is seen in emission only for the lowest surface gravities. As stated in Sect.\ \ref{s_sms_meth}, we adopted \logg\ = 0.8 in models A3 and A4 as an upper limit on the surface gravity. Given the trend shown in Fig.\ \ref{fig_t10L8}, we anticipate that models with lower \logg\ would show an even stronger emission across the Balmer break. This peculiar morphology is a key for identifying SMS in proto-GCs (see Sect.\ \ref{s_photom}). This feature is very important, we describe the details of its physics in the next section.

\subsubsection{Origin of the Balmer jump in emission in SMS}
\label{phy_bal}

\begin{figure}[t]
\centering
\includegraphics[width=0.49\textwidth]{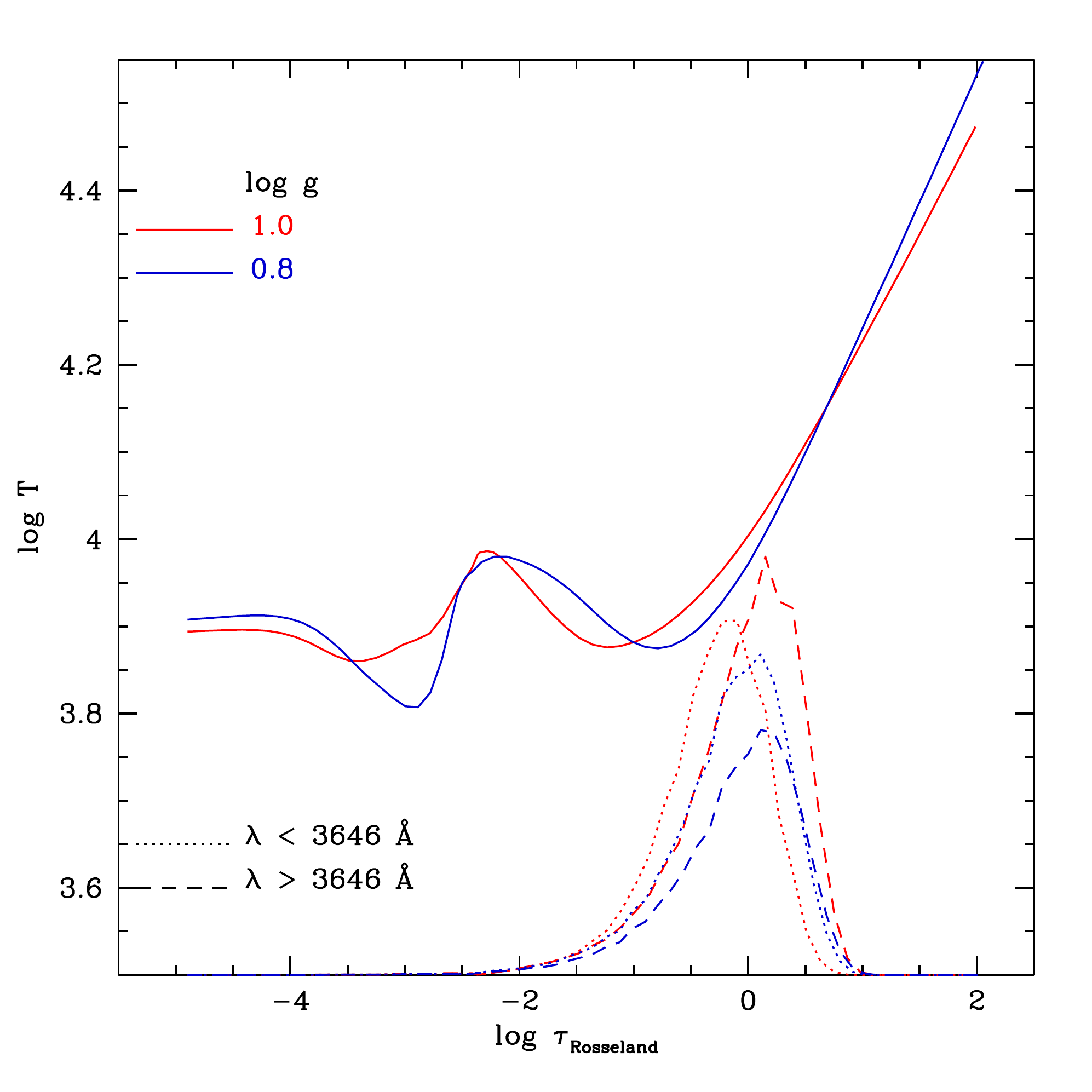}
\caption{Temperature structure of the models with \teff\ = 10000 K, \lL\ = 8.0 and \logg\ = 0.8 (blue) and 1.0 (red). The dotted and dashed lines show the contribution functions of the continuum blueward (3500 \AA) and redward (4200 \AA) of the Balmer jump, respectively.}
\label{fig_Tstru}
\end{figure}

\begin{figure*}[hbt]
\centering
\includegraphics[width=0.49\textwidth]{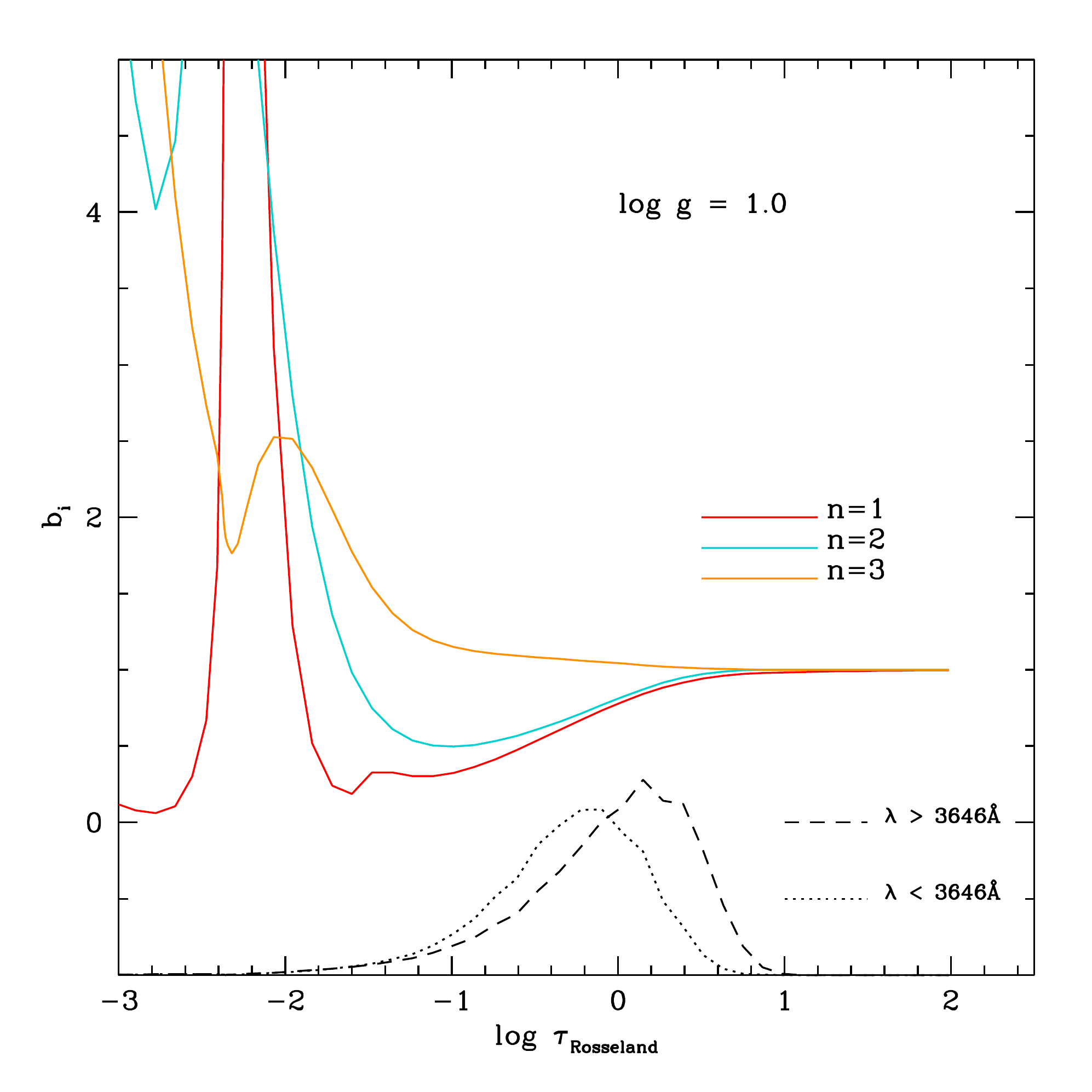}
\includegraphics[width=0.49\textwidth]{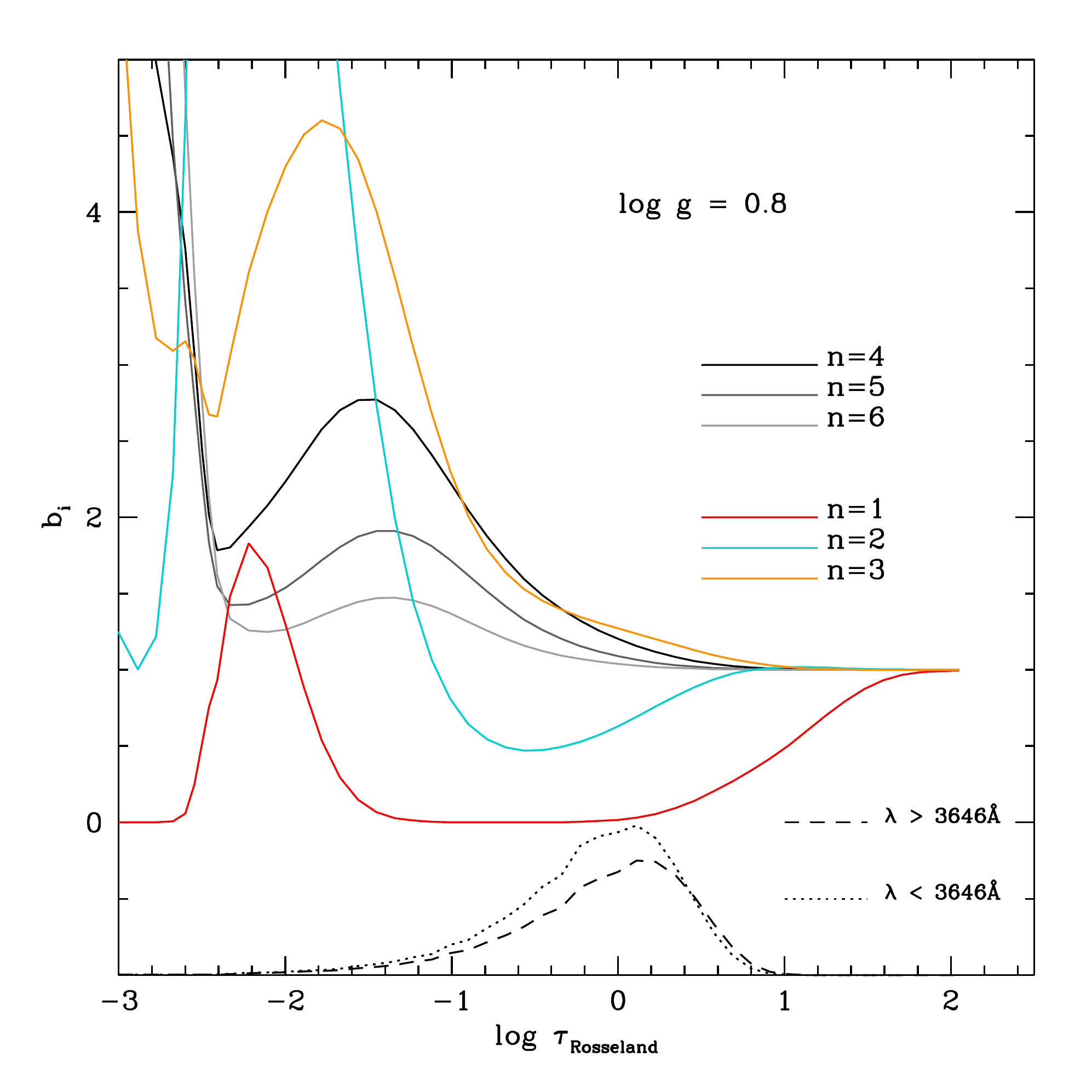}\\
\caption{Departure coefficients of levels 1, 2 and 3 in models with \teff\ = 10000 K, \lL\ = 8.0 and \logg\ = 0.8 (right) and 1.0 (left) as a function of Rosseland optical depth. The dotted and dashed black lines show the contribution functions of the continuum blueward and redward of the Balmer jump, respectively.}
\label{fig_bi}
\end{figure*}

Figures \ref{fig_Tstru} and \ref{fig_bi} show the temperature structure and the departure coefficients (see definition below), respectively, of the first Balmer levels in the two models with \teff\ = 10000~K and \lL\ = 8.0, but different \logg. Qualitatively, the models at \lL\ = 9.0 show the same behavior. The shape of the Balmer discontinuity can be understood as follows.

To first order, the specific intensity at the surface of the star in the radial direction is approximated by the Eddington-Barbier relation, 

\begin{equation}
\label{eq_barb}
I(0) = S(\tau_{\lambda} = 1)
,\end{equation}

\noindent where S is the source function and $\tau_{\lambda}$ is the optical depth at wavelength $\lambda$. Considering that the main source of opacity above and below the Balmer jump is continuum absorption from levels 3 and 2 of hydrogen, respectively, and following \citet{hm14}, we can approximate the source function by 

\begin{equation}
\label{eq_sb}
S_{\lambda} = \frac{1}{b_{i}} B_{\lambda}(T)
,\end{equation}    

\noindent where $b_{i}$ is the departure coefficient of level $i$, defined as the ratio of the level population to the population in LTE conditions.  $T$ is the temperature at an optical depth of 1.

Combining Eqs. \ref{eq_barb} and \ref{eq_sb} and assuming that the ratio of specific intensities reflects the ratio of emergent fluxes, we have

\begin{equation}
\label{eq_ratflux}
\frac{F_{3500}}{F_{4200}} = \frac{B_{3500}(\tau_{3500} = 1)}{B_{4200}(\tau_{4200} = 1)} \frac{b_{3}}{b_{2}}
,\end{equation}    

\noindent where we evaluated the flux ($F$) below (above) the Balmer jump at 3500 \AA\ (4200 \AA).

Fig.\ \ref{fig_Tstru} shows the temperature structure in the models whose spectra are displayed in Fig.\ \ref{fig_spec_Balmer}. The broken lines show the contribution function defined by \citet{hil87}. This quantity evaluates the fraction of the flux from different depths for a specific wavelength. We have chosen 3500 and 4200 \AA\ as representative wavelengths of the Balmer break. To first order, the maximum of the contribution function at wavelengths $\lambda$ is located at $\tau_{\lambda} = 1$. From Fig.\ \ref{fig_Tstru} we can estimate the temperature at the formation depth of the flux at wavelength $\lambda$. We can subsequently calculate the Planck function at this temperature at wavelength $\lambda$. For the model with \logg\ = 1.0 (0.8), we find that $\frac{B_{3500}(\tau_{3500} = 1)}{B_{4200}(\tau_{4200} = 1)} \sim 0.55 (1.2)$. In the \logg\ = 1.0 model, the flux at 4200 \AA\ is emitted from a region that is located deeper in the atmosphere, where the temperature is higher than in the formation region of the 3500 \AA\ flux. Therefore the ratio $\frac{B_{3500}(\tau_{3500} = 1)}{B_{4200}(\tau_{4200} = 1)}$ is lower than unity. For the \logg\ = 0.8 model, both continua are formed at the same depth, and the flux ratio is simply the ratio of the Planck function at two different wavelengths. It is thus higher than 1.0. 

To evaluate the second term in Eq.\ \ref{eq_ratflux}, we show in Fig.\ \ref{fig_bi} the departure coefficient in the two models. At the formation depth of the 3500 and 4200 \AA\ continua, we have $\frac{b_{3}}{b_{2}} = 1.4 (2.0)$ in the \logg\ = 1.0 (0.8) models. Combining these results with the blackbody flux ratios, we obtain $\frac{F_{3500}}{F_{4200}} = 0.8 (2.4)$ for the \logg\ = 1.0 (0.8) models. This means that the high (low) surface gravity model has a Balmer break in absorption (emission).

We note that according to Eq.\ \ref{eq_sb}, the departure coefficients are equal to 1 in LTE conditions. Following the Boltzmann equation, the ratio of the level n=2 to level n=1 populations is about 0.4 at 10000 K. We therefore expect a stronger opacity shortward of the Balmer break, which means that the continuum will be formed at a lower depth, in regions with a lower temperature. Based on Eq.\ \ref{eq_sb}, we therefore expect a Balmer break in absorption under LTE conditions.

\subsubsection{Hydrogen line spectra}

\begin{figure*}[t]
\centering
\includegraphics[width=0.49\textwidth]{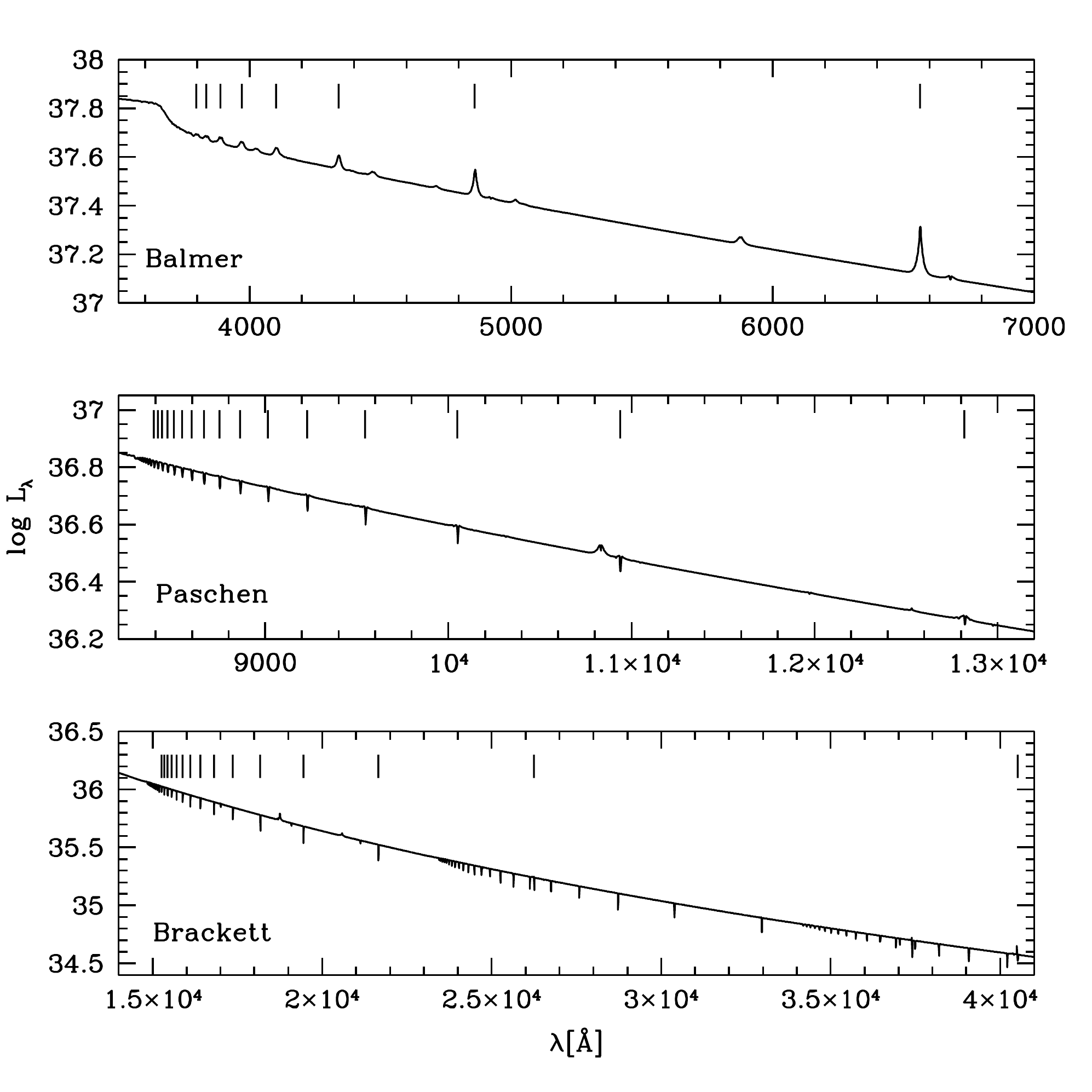}
\includegraphics[width=0.49\textwidth]{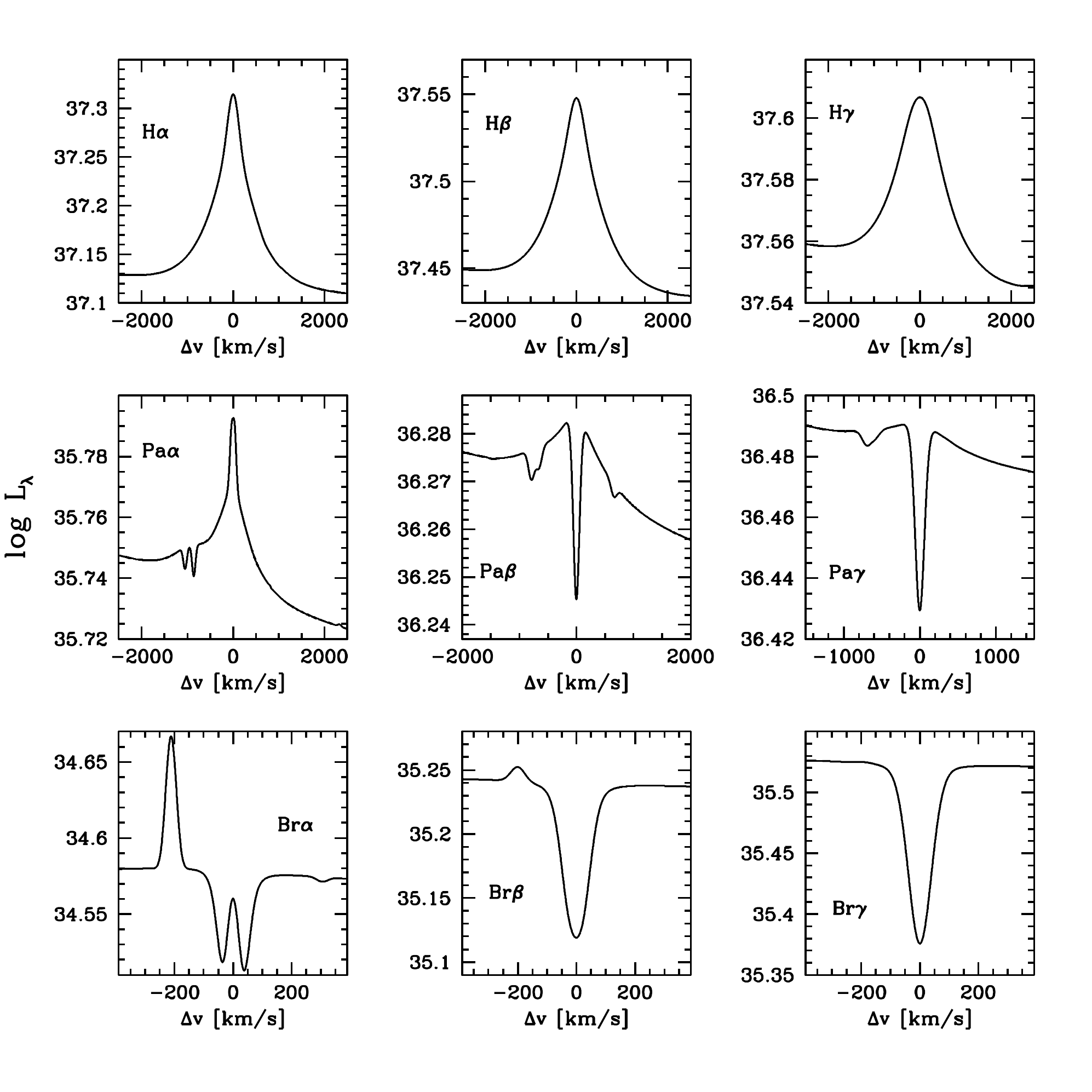}\\
\caption{\textit{Left panel}: Spectrum of model A3. The top (middle and bottom) panel shows the wavelength range encompassing the Balmer (Paschen and Brackett) lines. In each panel, the main lines of the series are indicated by vertical lines. \textit{Right panel}: Zoom into the first three lines of each series.}
\label{fig_Hseries}
\end{figure*}

\begin{figure}[t]
\centering
\includegraphics[width=0.49\textwidth]{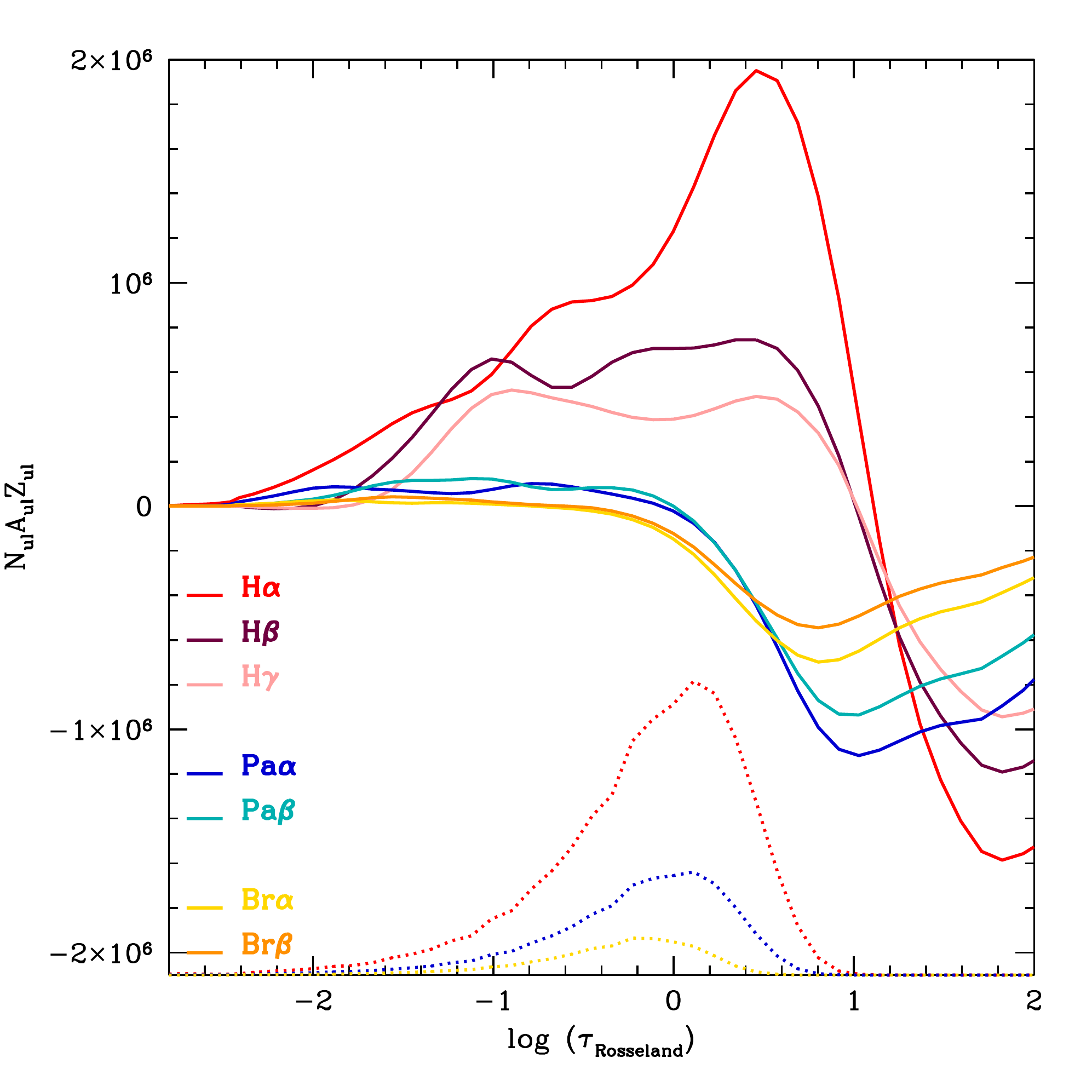}
\caption{Total radiative rate from the upper to the lower level in the hydrogen transitions of the Balmer, Paschen, and Brackett series as a function of Rosseland optical depth (solid lines) for model A3. The contribution function of H$\alpha$ (Pa$\alpha$ and Br$\alpha$) is shown by red (blue and yellow) dashed lines.}
\label{fig_rates}
\end{figure}

An interesting feature of the A3 and A4 models is that the lines of the Balmer series are in emission, while those of the other hydrogen series are in absorption, except for the Pa$\alpha$ (see Fig.\ \ref{fig_Hseries} for the \lL\ = 8.0 model, and Appendix \ref{ap_HlinesL9} for the \lL\ = 9.0 model). A quantitative explanation of this behavior is given in Fig.\ \ref{fig_rates}. In the strong non-LTE conditions encountered in the atmosphere of such models, radiative rates dominate widely over collisional rates. The total radiative rates (shown in Fig.\ \ref{fig_rates}) are positive when more photons are emitted than are absorbed in a given transition, and negative in the opposite case. In detailed balance, the rates are zero. In the formation region of the hydrogen lines (maximum of the contribution function) of model A3, the total rates are definitely positive in the Balmer lines, which explains that more photons are emitted than are absorbed, which leads to an emission line. Conversely, the radiative rates of the Brackett lines are negative, which explains the absorption lines. The rates are partly positive and partly negative for the Paschen lines. The line morphology is thus intermediate: it is in emission for Pa$\alpha$ and in absorption for the other members of the series. The emission in the Balmer lines can also be understood from the right panel in Fig.\ \ref{fig_bi}. In the line-formation regions (around $\tau_{\rm Rosseland} = 1.0$), the n=2 level (lower level of the Balmer lines) is depopulated (departure coefficient $\sim$ 0.5), while the n$\geq$3 levels are all overpopulated. There is thus a population inversion in all Balmer lines, which leads to emission. Table~\ref{tab_ew} gives the equivalent widths and luminosities of the H$\alpha$ and H$\beta$ lines in the models where at least one of the lines is predicted to be in emission. Another interesting feature is the quasi-absence of Lyman lines that is due to the extreme depopulation of the hydrogen ground level (see the right panel of Fig.\ \ref{fig_bi}).

In the models with \teff\ = 7000 K (A1 and A2), the hydrogen lines are all predicted in absorption (see Fig.\ \ref{fig_sed}, upper panels).

\begin{table}
\begin{center}
  \caption{Emission line properties.} \label{tab_ew}
\begin{tabular}{lccccc}
\hline
ID &  EW(H$\alpha$) &  log (L($\alpha$)/L$_{\odot}$)  & EW(H$\beta$) &  log (L($\beta$)/L$_{\odot}$) \\    
  K  &  [\AA]       &         &  [\AA]       &         \\
\hline
A3 &    11.1        &     5.62    &  5.1   &    5.79   \\
A4 &     8.7        &    6.62     &  4.2   &   6.80     \\
\hline
C1 &  1.8  &  2.98   &  --    &   --  \\
\hline                                                                     
\end{tabular}                                                              
\tablefoot{The wavelength ranges in which the equivalent widths and luminosities are calculated are 4820-4900 \AA\ for H$\beta$ and 6510-6620 \AA\ for H$_{\alpha}$. For model C1, H$\beta$ is an absorption line.}
\end{center}                                                     
\end{table}

\subsection{Intermediate temperature and hot SMS} 

Models in which \teff\ is higher than 10000 K are presented in the middle and bottom panels of Fig.2. The hottest models (B1 and C1) have an emission peak in the far-UV and an almost featureless spectrum. The optical flux is much lower than in models with lower temperature. 

\begin{figure}[t]
\centering
\includegraphics[width=0.49\textwidth]{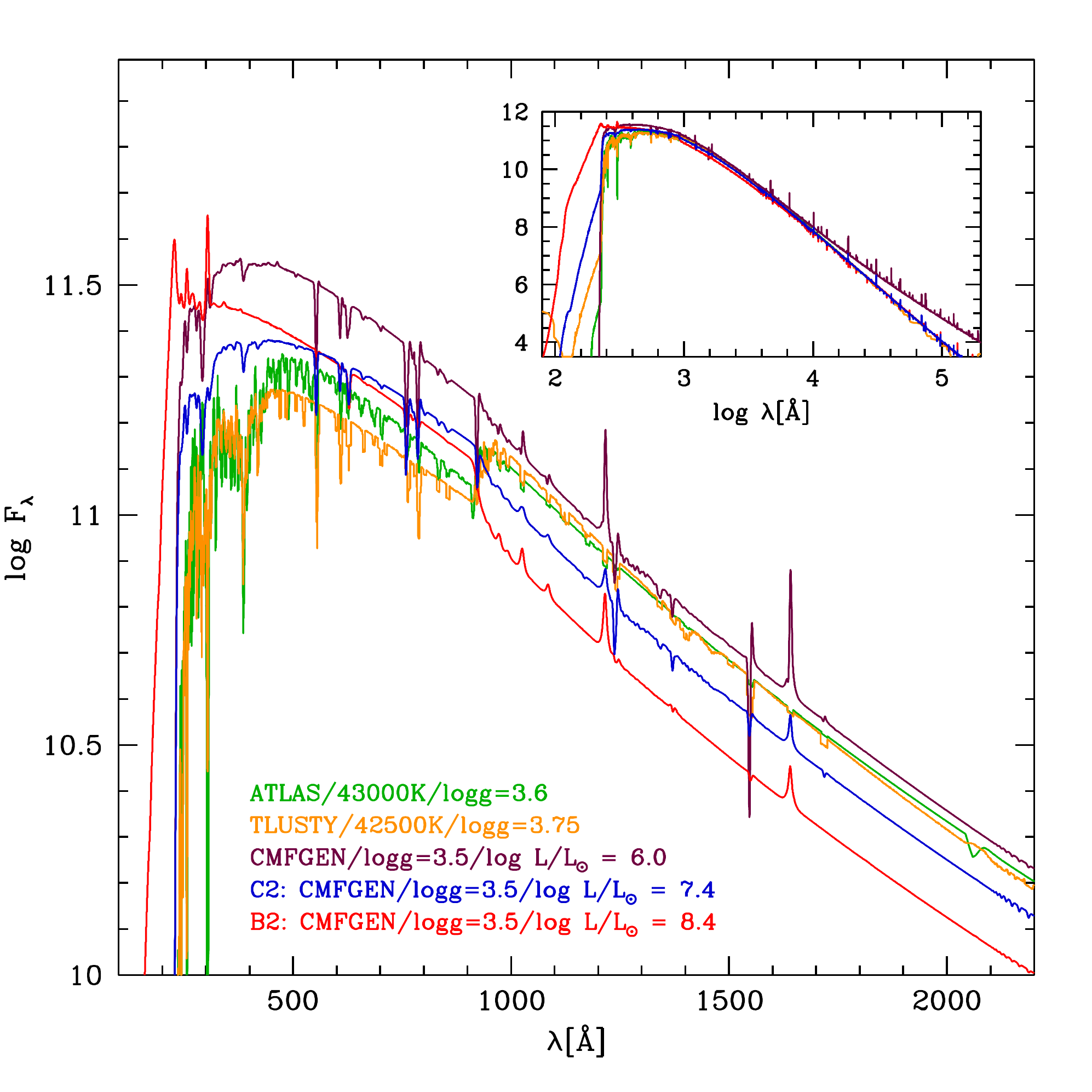}
\caption{Comparison between the SED of our model B2 (red) and C2 (blue) and 1) an \emph{ATLAS} model with the same \teff\ and a slightly higher \logg\ (green), 2) a \emph{TLUSTY} model with \teff\ = 42500 K / \logg\ = 3.75 (orange), and 3) a \emph{CMFGEN} model with luminosity reduced to \lL\ = 6.0 (dark red). The fluxes are given in erg/s/\AA/cm$^2$.}
\label{fig_t43}
\end{figure}

The morphology of the spectra in models with the highest effective temperatures (higher than 43000 K) can be compared to that of the models at 10000 K. The Lyman break in the hot models is indeed observed in emission (see Figs.\ \ref{fig_sed} and \ref{fig_t43}). The Lyman lines are also in emission, whereas the lines from higher order series are in absorption. In analogy to the explanation of the Balmer break morphology for models A2 and A4 , we argue that the same non-LTE effects are at work for the Lyman break. In the following, we focus on models with \teff\ = 43000 K and take them as representative of "hot" SMS.

Fig.\ \ref{fig_t43} compares models B2 and C2 to other models with similar \teff. The \emph{ATLAS} model has a Lyman break in absorption and a smaller flux shortward of it than do models B2 and C2. Above the Lyman break, the situation reverses and the \emph{ATLAS} model emits more flux. This highlights once more the need for non-LTE models for SMSs.
When the luminosity increases in the \textit{CMFGEN} models (from a standard value for normal massive stars -- \lL\ = 6.0 -- to those of SMSs), the Lyman break shows a stronger emission morphology. The far-UV flux increases, and consequently, the optical flux decreases (relatively). The non-LTE effects are stronger for the extreme luminosity conditions encountered in SMS atmospheres. It can be shown (see Appendix\ \ref{ap_hot}) that in the most luminous model (B2), the continuum shortward of the Lyman break is formed closer to the photosphere than in model C2, while the continuum longward of it is formed in the same region in both models. The departure coefficients of the ground and first excited levels are similar (~0.7-0.8 vs. ~1.1-1.4) in the formation regions of the continua. The temperature effect in Eq.\ \ref{eq_sb} dominates and explains that a larger jump across the Lyman break is observed in model B2. The additional \emph{CMFGEN} model in Fig.\ \ref{fig_t43}, corresponding to a normal massive star (M$\sim$100 \msun), does not show any sudden change across the Lyman break (in emission or in absorption). In comparison, the \emph{TLUSTY} model shows a strong Lyman break in absorption. This highlights the importance of taking not only non-LTE effects into account, but also wind effects (\emph{TLUSTY} models are plane parallel). 

Another interesting feature of models B2 and C2 is the behavior at long wavelengths (above $\sim$1 \mum). The model for a normal massive star (\lL\ = 6.0) shows a well-known excess of emission due to free-free processes in the extended atmosphere. This excess is almost absent in models B2 and C2 because of their low wind density (see Sect.\ \ref{ap_hot} for a detailed explanation). Consequently, the difference between models that include winds and plane-parallel models is small at long wavelenghts.

\section{Predictions for proto-GCs hosting SMS at high redshift}
\label{s_starclu}

We now proceed to simulate spectra and synthetic photometry of SMS and their host proto-GCs at high redshift. This will allow us to predict their basic observable properties and to examine possible signatures that allow us to distinguish ``normal'' young massive clusters from those hosting SMSs.

\subsection{Method}

Emission from proto-GCs hosting SMS contains the following contributions, which must be taken into account: emission from the SMS, the young stellar cluster hosting the SMS, and nebular emission (continuum and lines) from the ionized gas surrounding the young cluster. The SMS spectra have been discussed above (Sect.\ \ref{s_sms_prop}). The physical parameters of the SMS and the surrounding cluster are connected and are described by the model of \cite{gieles18}, which in particular relates the cluster mass to the mass and radius of the SMS for a given value of the parameter $\delta$.

We primarily considered the case of very massive ($\mstar \ga 10^4$ \msun) and cool SMS, with \teff $\sim 7000-14000$ K, which corresponds to the case of $\delta \approx 1$ of  \cite{gieles18}. This situation is probably the most likely because SMSs should have a high accretion rate (hence a large radius and low \teff) most of the time. It should also be the most favorable case where an SMS surrounded by the young cluster can be distinguished observationally, as we show below, in contrast to a young cluster with a compact and hot SMS. In addition, the case where $\delta \approx 1$ corresponds to a high-mass (and thus to first-order luminosity) ratio between the SMS and the surrounding cluster (see Fig.~3 of \cite{gieles18}), which maximizes the detectability of the SMS. 

We computed the following three cases that are summarized in Table \ref{tab_clusters}: 
{\em 1)} SMS model A2 + cluster, {\em 2)} SMS model A4 + cluster, and {\em 3)} SMS model B3 + cluster. These cases all correspond to cases of $\delta \sim 1$. From Fig.\ 3 of \cite{gieles18} we can then read off the cluster mass, which we adopt as $M_{\rm cl} = 5 \times 10^5$ \msun\ (number of stars $N \sim 10^6$), with a Kroupa-like IMF from 0.1 to 100 \msun\ \citep[cf.][]{gieles18}. 

The cluster spectrum was taken from the synthesis models of \cite{2003A&A...397..527S} for 1/50 solar metallicity and an age of 2 Myr, corresponding to $t \sim 4$ Myr on the timing of gas accretion in \cite{gieles18}. The cluster SED does not strongly depend on metallicity and is very similar to that obtained from other synthesis models \citep[e.g.,][]{1999ApJS..123....3L,2003MNRAS.344.1000B}, which we also used for comparison purpose (see Sect.\ \ref{s_compother}).

We then used a modified version of the photometric redshift and SED fitting code {\em Hyperz} \citep{schaerer&debarros2009} to synthesize observables for the summed SMS + cluster SEDs, allowing also for the contribution of nebular emission, which is proportional to the amount of Lyman-continuum photons that are emitted (from the normal hot stars in the cluster).
We adopted the mean IGM transmission of \cite{madau95} and allowed for dust attenuation in some cases, using the Calzetti law \citep{calzettietal2000}. 
Finally, we adopted standard cosmological parameters: $H_0 = 70$ km s$^{-1}$ Mpc$^{-1}$, $\Omega_\Lambda = 0.7$, and $\Omega_m = 0.3$. 

We predict a synthetic photometry for a range of broadband filters that are offered with the NIRCam and MIRI instruments on board the {\em JWST} that cover the wavelength range from $\sim 0.7$ to 8 micron. This includes the following filters:
{\em F070W}, {\em F090W}, {\em F115W}, {\em F150W}, {\em F200W}, {\em F277W}, {\em F356W}, {\em F410W}, and {\em F444W} from NIRCam, plus {\em F560W} and {\em F770W} from MIRI\footnote{See \url{https://jwst-docs.stsci.edu/} for documentation}.
The tables in which the results are listed are described in Appendix \ref{ap_photom} and are available at the CDS.

\begin{table}
  \caption{Main parameters of the proto-GCs (cl) + cool SMS ($\delta \approx 1$) spectral models we computed} 
  \label{tab_clusters}
\begin{tabular}{lccccc}
\hline
SMS & $M$/\msun & $L/\lsun$  & \teff\ [K] &  $M_{\rm cl}/\msun$  & $L_{\rm cl}/\lsun$ \\
\hline
A2 &  $2.5 \times 10^4$ & $10^{9.0}$ & 7000   & $5 \times 10^5$ & $10^{9.0}$ \\
A4 &  $5 \times 10^4$    & $10^{9.0}$ & 10000 & $5 \times 10^5$ & $10^{9.0}$ \\
B3 &  $1 \times 10^4$    & $10^{8.4}$ & 13600 &  $5 \times 10^5$  & $10^{9.0}$\\
 \hline                                                                   
\end{tabular}                                                            
\end{table}

\begin{figure*}[htb]
\centering
\includegraphics[width=0.49\textwidth]{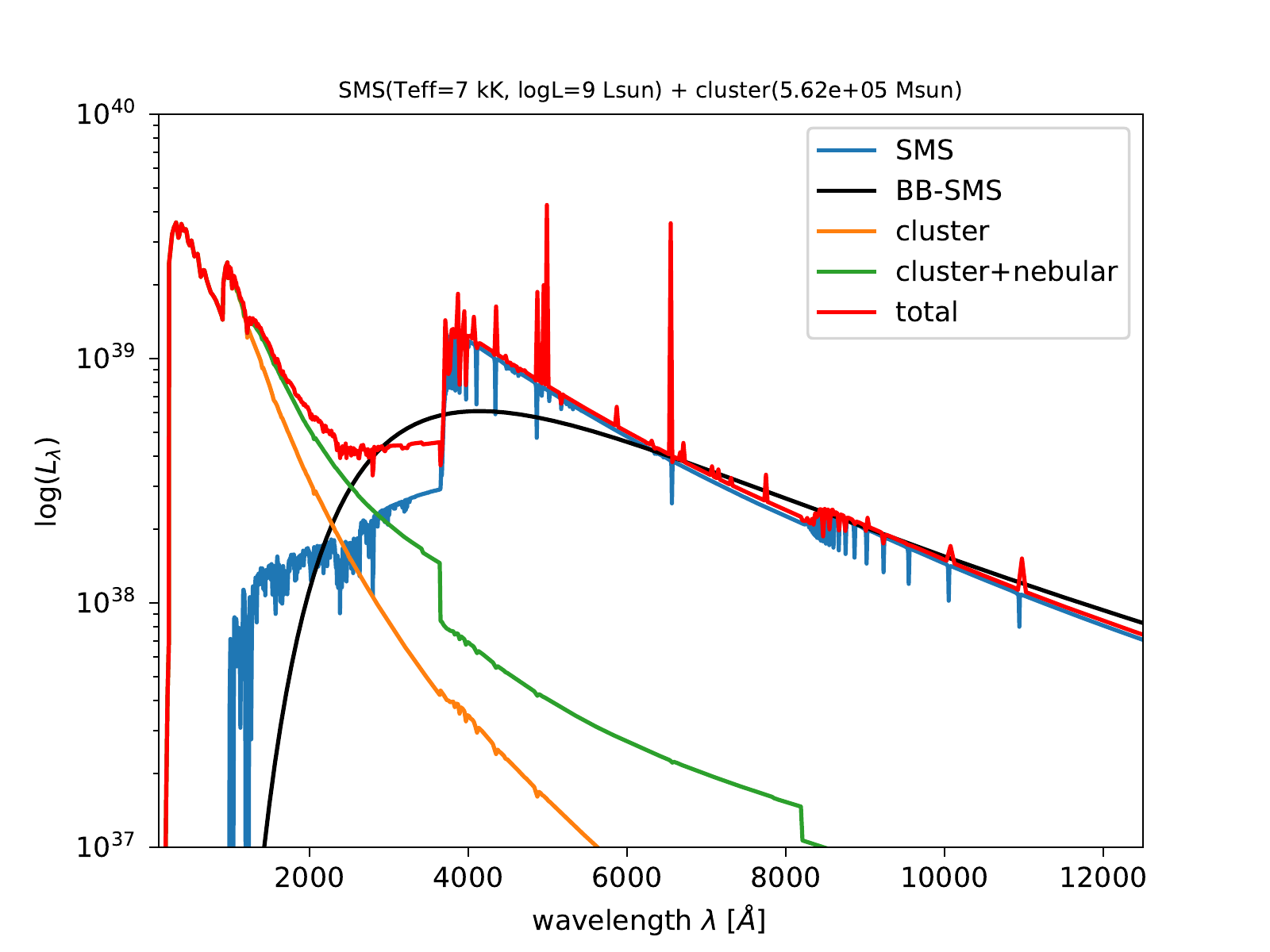}
\includegraphics[width=0.49\textwidth]{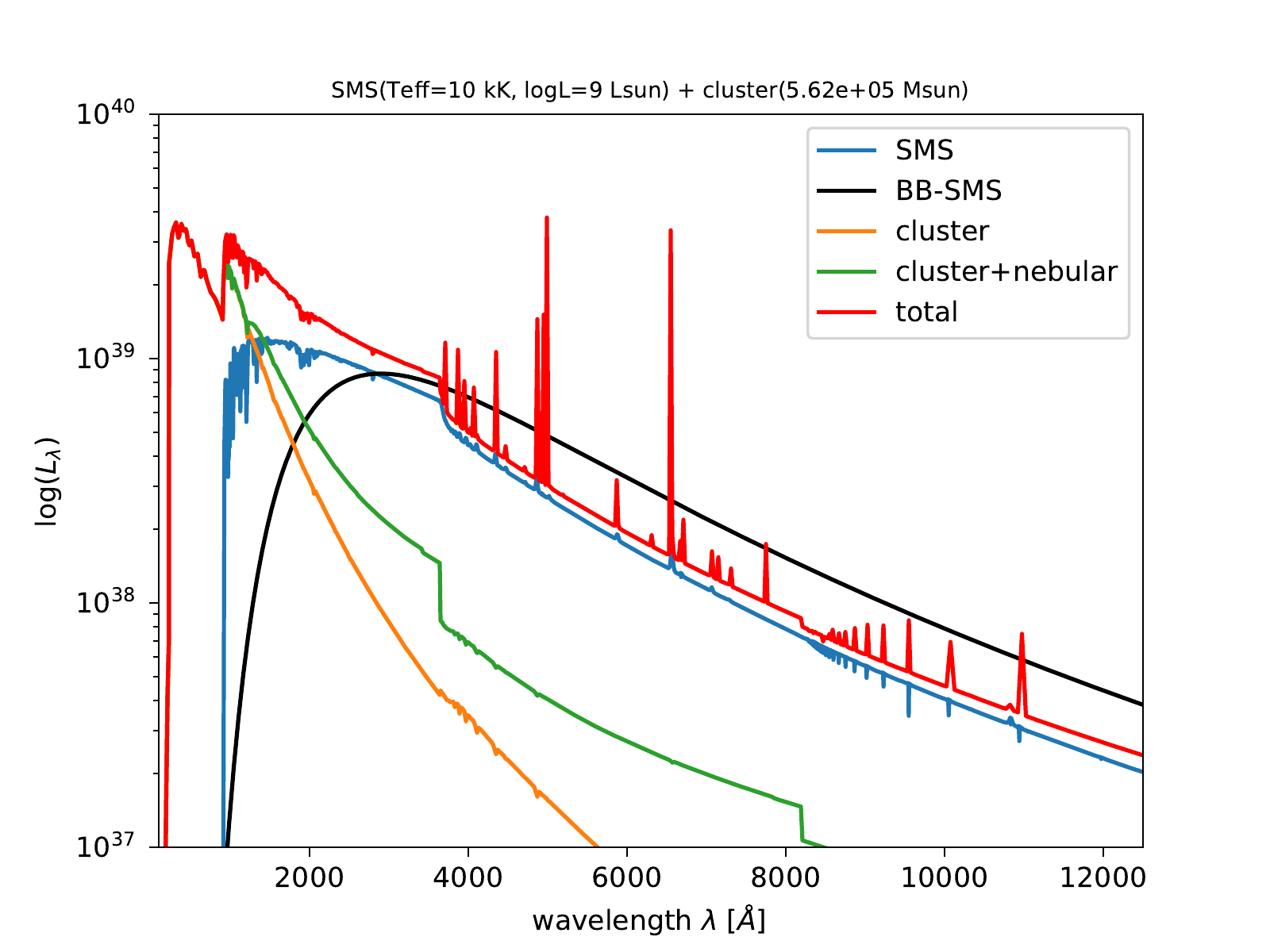}
\caption{Spectral energy distributions of  proto-GCs containing a cool SMS (blue lines show model A2 on the left, and model A4 on the right) plus a normal young stellar population (orange). The green curve indicates the nebular continuum emission from the young population. The total spectrum, also including nebular emission lines, is shown in red. The black curve shows a blackbody with the same \teff\ and $L$ as the SMS. The clear dominance of the SMS spectrum at wavelengths $\lambda \protect\ga 3000$ (1250) \protect\AA\ for model A2 (A4) is obvious.}
\label{fig_A24}
\end{figure*}

\begin{figure}[htb]
\centering
\includegraphics[width=9.8cm]{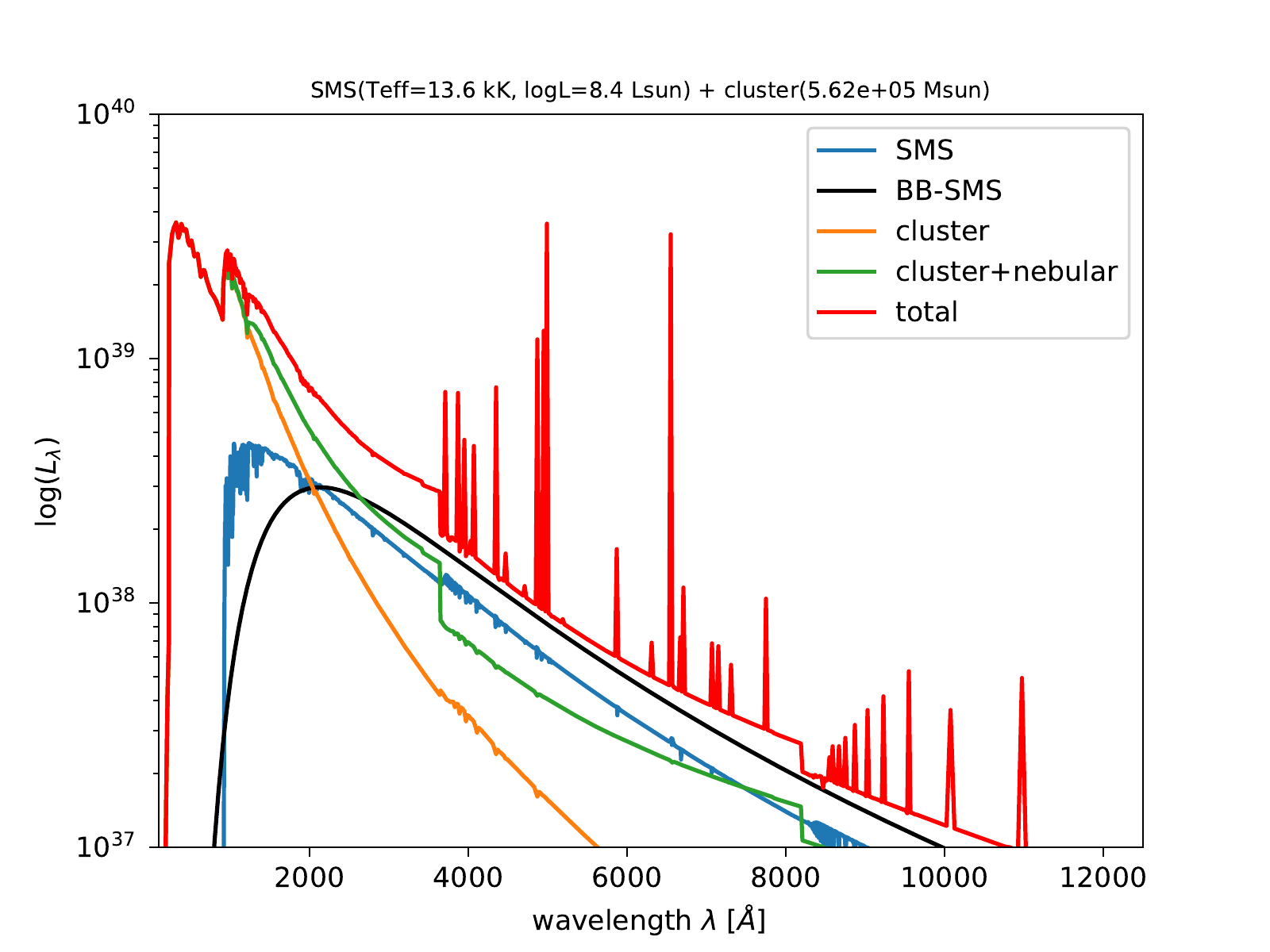}
\caption{Same as Fig.\ \protect\ref{fig_A24}, but for the less luminous and slightly hotter SMS B3. Here the emission from SMS is subdominant in the UV and comparable to the nebular emission from the cluster in the optical range.}
\label{fig_B3}
\end{figure}

\subsection{Predicted SEDs}
\label{s_seds}

The resulting SEDs of the three cases are shown in Figs.\ \ref{fig_A24} and \ref{fig_B3}.
The most important result is that the fluxes of SMS with luminosities $\log L \sim 8.4-9.0$ \lsun\ are comparable to or higher than the flux of the surrounding young stellar cluster (which has a total luminosity of $\log L \sim 9.0$ \lsun).
For the brightest SMS ($\log(L/\lsun)=9$), the SED is entirely dominated by emission from the SMS at wavelengths $\ga 1200$ \AA\ (model A4) and $\ga 3000$ \AA\ for the coldest SMS (model A2). The emission of the young stellar cluster peaks in the UV (and provides all the Lyman-continuum radiation), whereas the SMS dominates at longer wavelengths, even when nebular continuum emission from the \hii\ region is taken into account.
These SEDs of the proto-GC in the SMS phase clearly show a fairly unusual shape that is not comparable to that of normal SSPs, but which we would characterize as 
``composite'' SEDs, resembling the superposition of a young and an old population.

\begin{figure}[htb]
\centering
\includegraphics[width=0.49\textwidth]{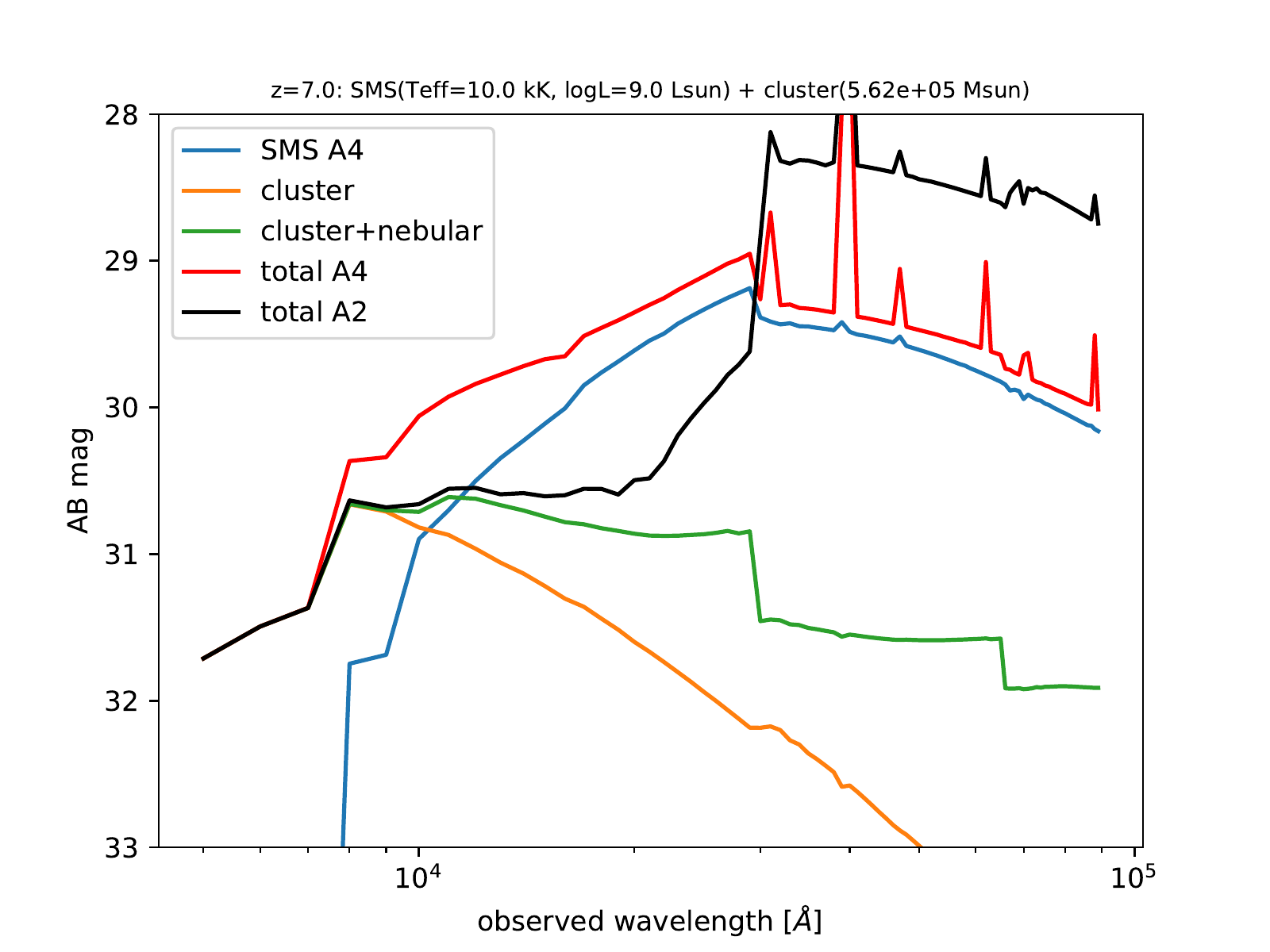}
\caption{Observed SED in AB magnitudes showing the proto-GC with two models for the SMS, A2 and A4 (shown in Fig.\ \protect\ref{fig_A24}), seen at redshift $z=7$.}
\label{fig_z7}
\end{figure}

Figure \ref{fig_z7} illustrates the predicted spectrum of these two cases with a proto-GC hosting the bright SMS described by models A2 and A4, as observed in situ during its formation at high redshift (here $z=7$).
The predicted magnitudes are about \magab $\sim 28.5-30.5$ from $\sim$ 0.8 to 10 \mum,  depending on the model. In the rest-UV the brighter model (A2) has an absolute UV magnitude of $M_{\rm UV} \sim -17.5$.
Between Ly$\alpha$ and $\sim 2500$ \AA\ in the rest-frame, the total spectrum shows a UV slope $\beta \sim -2$ (i.e.,\ close to constant in AB magnitude) for model A2 and a much redder slope of  $\beta \sim -0.8$ for A4.

The SEDs again have an unusual shape, with fairly red rest-UV spectra (observed at $\lambda \la 3$ micron) and then a bluer slope after the flux peak of the SMS in AB magnitudes ($\lambda > 3$ \mum). For the coolest SMS (model A2), the flux increase from the rest-UV across the Balmer break to the rest-optical is quite strong ($\sim 1.5-2$ mag). Clearly, this peculiarity of the SED can be used as a distinction, as shown in more detail below (Sect.\ \ref{s_photom}).

The SED of the proto-GC + SMS B3 shown in Fig.\ \ref{fig_B3} illustrates that hot SMS are not distinguishable from the cluster population. SMS model B3 barely dominates the light budget in the optical. Hotter models have either lower masses or smaller $\delta$ (see Table \ref{tab_param}). In both cases, the contribution of the cluster to the total flux should be larger than the case shown in Fig.\ \ref{fig_B3} (see also Sect.\ \ref{s_discuss}). Consequently, the SMS would remain undetected. In the remainder of Sect.\ \ref{s_starclu} we thus focus on the three cluster+SMS combinations shown in Figs.\ \ref{fig_A24} and \ref{fig_B3}.

\subsection{Synthetic photometry and distinct observational features of SMS}
\label{s_photom}

\begin{figure*}[htb]
\centering

\includegraphics[width=0.49\textwidth]{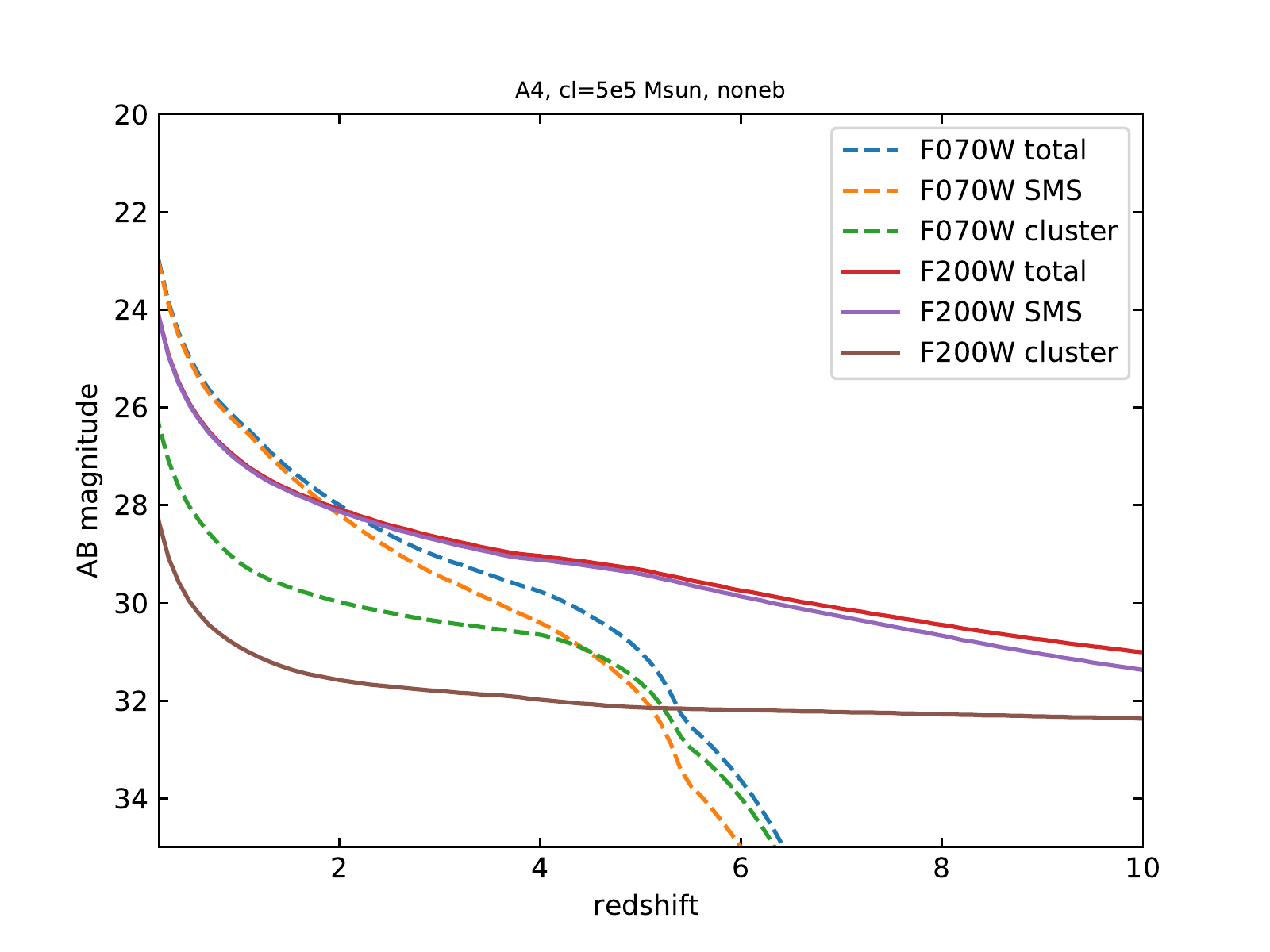}
\includegraphics[width=0.49\textwidth]{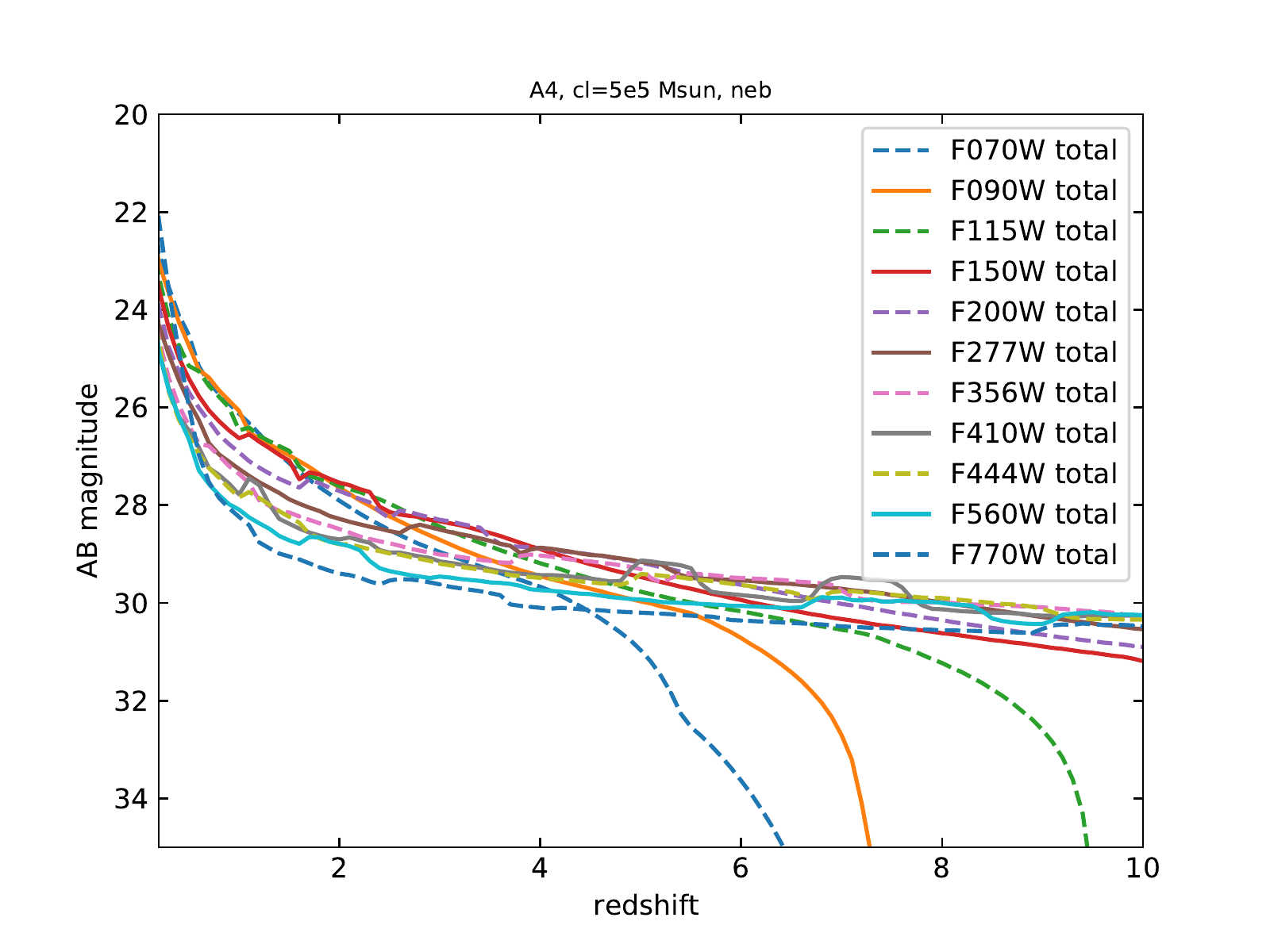}
\caption{{\em Left:} Predicted F070W and F200W magnitudes for model A4 as a function of redshift showing the contributions of the stellar cluster (without nebular emission), the SMS, and the total spectrum. {\em Right:} Predicted magnitudes in the selected NIRCAM+MIRI filters
from 0.7-7.7 micron for the same proto-GC + SMS model. The small fluctuations in the different bands are due to nebular emission features. The three curves that drop out are due to the classical photometric drop-out from the shortest wavelengths filters.}
\label{fig_mags_z}
\end{figure*}

In Fig. \ref{fig_mags_z} (left) we show the predicted magnitude of the proto-GC+SMS for model A4 in broadband filters at 0.7 and 2 \mum\ up to $z=10$. 
Over a wide redshift range ($z>2$), the total magnitude is between \magab$=28$ and 30.  At lower redshift, brighter magnitudes up to $\sim 24-26$ are predicted.
This figure also shows the significant gain in brightness (by up to $\sim 3$ mags in the 2 \mum\ filter) that is due to the SMS, which strongly dominates the rest-frame optical emission that is probed by this filter.
At shorter wavelength, for example, at the wavelength\ probed by the 0.7 \mum\ filter, the difference is smaller, as expected.

The right panel of Fig. \ref{fig_mags_z} shows the predicted magnitudes in all the synthesized filters from 0.7 to 7.7 \mum\ of the same model (cluster+A4). With its SED shape, dominated by a cool (10 kK) SMS
showing an increased UV flux (due to non-LTE effects),  a relatively constant AB magnitude is expected at these wavelengths with \magab=$28-30$ at $z \ga 3$, except at the shortest wavelengths, where the flux drops due to IGM attenuation.
The behavior is quite similar for model A2, although it is somewhat fainter at the shortest wavelengths (see, e.g.,\ Fig.\ \ref{fig_z7}).
In passing we note that the magnitudes predicted for proto-GCs with bright SMS shown here are significantly brighter than those predicted for the typical GC precursors discussed by \cite{Pozzetti2019Search-instruct}, although they assumed a cluster mass that was four times higher. This difference is obviously primarily due to the strong emission from the SMS, and after the SMS phase, our predictions will again resemble those of normal SSPs.

\begin{figure}[tb]
\centering
\includegraphics[width=0.49\textwidth]{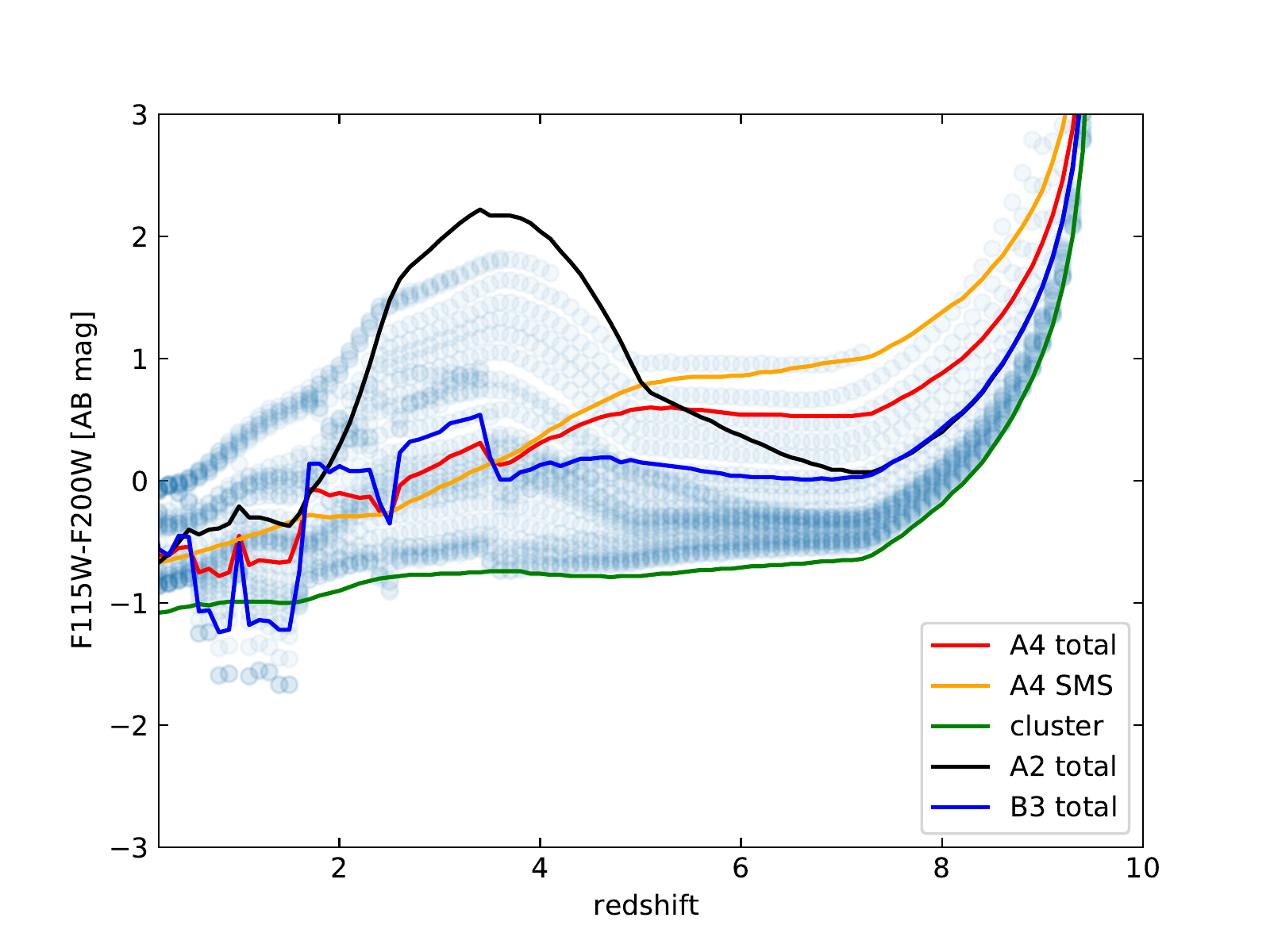}
\caption{Color-redshift plot showing the synthetic F115W-F200W color of the three proto-GC+SMS models and the colors of the different components. The total SEDs are shown with nebular emission, and the cluster model is shown without it. The underlying blue-gray points show the synthetic color of SSPs (clusters) of ages from zero to the maximum allowed age at each redshift.
}
\label{fig_col_z}
\end{figure}

\begin{figure}[tb]
\centering
\includegraphics[width=0.49\textwidth]{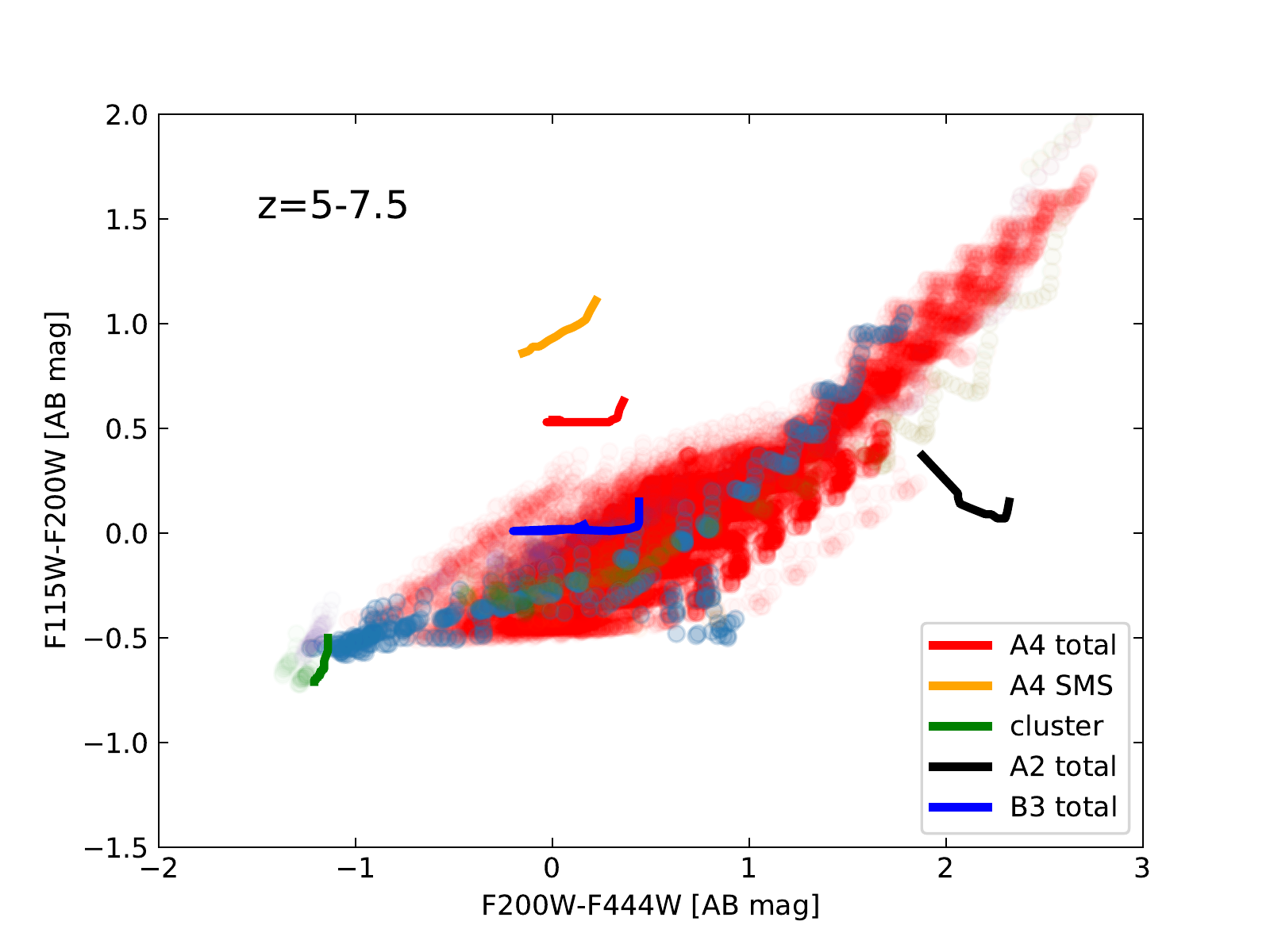}
\caption{Synthetic  (F115-F200W) vs. (F200W-F444W) color-color plot for the three proto-GC+SMS models (colored lines, same models as in Fig.~\protect\ref{fig_col_z}) at redshift$z=5-7.5$. The underlying colored cloud shows the location of normal SSPs and integrated populations with exponentially decreasing star formation. The simulations also allow for reddening with $A_V$ up to 1, which introduces an additional dispersion along the diagonal in the diagram. }
\label{fig_col_col}
\end{figure}

The question now is whether the colors of young proto-GCs with SMS can be distinguished by photometry from
normal clusters or other objects. 
Figure \ref{fig_col_z} shows as an example the (F115W-F200W), that is, the (1.5-2 \mum) color, of the three models compared to the color of the hosting young cluster, SSPs with different ages, and populations with exponentially declining star formation histories.
At each redshift, the youngest clusters have the bluest (F115W-F200W) color, which then becomes redder with increasing age. The colors of the proto-GCs with model A2 and A4 are both significantly redder than that of the surrounding young cluster, but model B3 is less distinct. In the redshift range $z \sim 2.5-4.5,$ the A2 model shows a very red (F115W-F200W) color because its uniquely strong Balmer break passes through these filters. This is the case because as a result of dilution from other stars, any normal stellar population has a smaller Balmer break than the strongest break of a single star.
This, or other filter combinations probing the Balmer break of the SMS, might be used to detect sources that are dominated by very cold and luminous SMS. For less extreme SMS spectra, single colors can probably not be used to uniquely determine their presence, and color-color plots or similar methods should be used. 

As shown in Fig.\ \ref{fig_col_col}, proto-GCs with cool and bright SMS show distinctive features in some appropriate color-color diagrams. For example, when we select two filters in the rest-UV range (shortward of the Balmer break) and a third filter longward of the Balmer break, we can measure the peculiar shapes of the SMS-dominated spectra, as shown\ in Fig.\ \ref{fig_z7}, for instance.
At redshifts $z \sim 5$ to 7.5, this can be done using the {\em F115W}, {\em F220W}, and {\em F444W} NIRCAM filters, which have central wavelengths $\sim$ 1.15, 2.2, and 4.4 \mum,\ respectively.
The cool+bright SMS-dominated SED A4 clearly stands out with a significantly redder (F115W-F200W) color
at (F200W-F444W) $\sim -0.2$ to 0.4 compared to the colors predicted for normal stellar populations\footnote{The models shown here include not only SSPs, but also populations with exponentially declining star-formation histories with different e-folding timescales. We also allow for reddening with $A_V$ up to 1 with the Calzetti law \citep{calzettietal2000}.}, shown here by the colored background shape. The reason is evident in Fig.\ \ref{fig_z7} and has been discussed there (Sect.\ \ref{s_seds}).
The coolest SMS (model A2) with the strong Balmer break moves away from the region of normal stellar populations toward much redder (F200W-F444W) for redshifts $z \ga 6$ when this color  properly measures the Balmer break. The spectra of hotter models (B3 or even hotter SMS) become indistinguishable from normal stellar populations, as is also shown by Fig.\ \ref{fig_col_col}. 

In summary, even with an approximate knowledge of the source redshifts (that can, e.g.,\ be derived from the classic photometric redshift method, which at high-$z$ relies on the Lyman (or Lyman-$\alpha$) break), two color measurements appropriately selected shortward and across the Balmer break can uniquely identify proto-GCs with cool (\teff $\la 10000$ K) and luminous SMS. 

\subsection{Other observables}
\label{s_linesclu}

In addition to the overall spectral shape just discussed, other features may also be somewhat peculiar in the scenario explored here,
where a normal stellar population with a supposedly normal maximum mass (here assumed to be $\sim 100$ \msun) co-exists with an SMS for some time. As shown in Sect.\ \ref{s_sms_prop}, the spectra of the SMS generally show H lines in absorption or weakly in emission, with equivalent widths not exceeding $\sim 11$ \AA\ for \ha\ (see Table \ref{tab_ew}). In contrast, the young ($\sim 2-4$ Myr) stellar population surrounding the SMS will create an \hii\ region that gives rise to strong hydrogen recombination lines plus the usual metal lines, whose strength will primarily depend on the gas metallicity and ionization parameter. The \ha\ equivalent width of such a population exceeds 500 \AA\ and can reach up to EW(\ha)$\sim 2000-3000$ \AA\ \citep[cf.][]{1999ApJS..123....3L}.
Because the continuum at \ha\ wavelengths is dominated by the SMS, whose flux can be $\sim 8-10$ times higher than that of the cluster
(including nebular continuum; cf.\ Fig.\ \ref{fig_A24}), the \ha\ emission with equivalent widths is expected to be not higher than EW(\ha)$\sim 200-400$ \AA\ from the \hii\ region, plus some broad but faint emission component from the SMS or underlying absorption. However, given the faintness of the line flux from the SMS (see Table \ref{tab_ew}), this component will probably remain undetectable at high redshift.
For the assumed mass and age of the proto-GC ($5 \times 10^5$ \msun\ and 2 Myr), the \ha\ luminosity is on the order of $L(\ha)=3.3 \times 10^{40}$ \ergs, which at $z=7$ corresponds to a flux of $F(\ha)=5.7 \times 10^{-20}$ \ergscm. Such faint fluxes are difficult to measure spectroscopically, even with the NIRSPEC spectrograph on board the JWST.

Emission lines such as \lya\ and weaker metal lines (e.g.,\ C~{\sc iv} $\lambda$1550, C~{\sc iii}] $\lambda$1909), known to be present in UV rest-frame spectra of metal-poor \hii\ regions and star-forming galaxies, could also be detected in spectra of proto-GCs.
For proto-GCs with cold SMS,  the strengths (EWs) of the UV lines (originating from the normal cluster population) will be significantly less altered than those of the optical lines. For model A4, for example, the continuum flux at $\sim 1215$ \AA\ is approximately the same as from the  cluster + SMS, which implies that the \lya\ EW will be reduced by a factor $\sim 2$ in presence of this SMS. For SMS models A2 and B3, \lya\ will be less strongly modified. In short, because young clusters emit strong \lya\ with EWs up to $\sim 200-300$ \AA\ \citep[cf.][]{2003A&A...397..527S}, we also expect strong intrinsic \lya\ emission from proto-GCs.
In case B, the \lya\ flux is $F(\lya)=4.6 \times 10^{-19}$ \ergscm\ for the same proto-GC mass as discussed above. This is approximately twice lower than the faintest \lya\ line fluxes measured with MUSE up to $z \sim 5.5-6.5$ \citep{Drake2017The-MUSE-Hubble}, but well within the range of the upcoming generation of 30m class telescopes \citep[see, e.g.,][]{Evans2015}.

At longer wavelengths, the continuum of the SMS becomes stronger; therefore the equivalent widths of other UV emission lines (e.g.,\ C~{\sc iii}] $\lambda$1909, O~{\sc iii}] $\lambda$1666) will be more reduced than \lya. In any case, these lines are generally fainter than \lya  \ and are therefore even more difficult to detect.
Finally, for the lack of S/N, it currently seems out of reach to measure the UV absorption features (e.g.,\ at $\sim 1600$ and $\sim 2400-2600$ \AA\ primarily due to large numbers of iron lines) predicted in the spectra of the cool SMS stars at large cosmological distances.

In short, detecting emission or absorption lines from proto-GCs with cool SMS is probably not feasible with current instruments, except for gravitationally lensed systems. \lya\ is the only emission line that may be observed with the new-generation observatories.

\section{Discussion}
\label{s_discuss}

\subsection{Variations of observable properties with proto-GC/SMS mass}
\label{s_mass}

In the scenario of GC formation examined here, the mass of the stellar cluster and that of the SMS grow in lockstep 
after the initial gas accretion phase and start of the runaway collision phase. To first order, $M_{\rm SMS}$ then scales linearly with cluster mass $M_{\rm cl}$ during a phase that approximately lasts for 3 Myr \citep[see Fig.\ 3 of ][]{gieles18}.
The exact mass scale of the SMS depends on its growth through collisions, which is more effective when the SMS is
bloated, that is, when it\ has a large radius, which is parameterized with $\delta$ through a mass-radius relation. For values of $\delta \approx 1,$
the ratio $M_{\rm SMS}/M_{\rm cl}$ is highest, hence also $L_{\rm SMS}/L_{\rm cl}$, which means that this will maximize 
the contrast between the SMS and the cluster SED. These cases have been presented in detail above (Section \ref{s_starclu} and
Table \ref{tab_clusters}). 
For smaller radii ($\delta=0.5$), the same SMS mass is reached only for clusters that are approximately ten times more massive \citep[Fig.\ 3  of ][]{gieles18}, which means that they are\ also more luminous by the same amount. In these cases the SMS fraction is clearly too low in both stellar mass and luminosity to significantly contribute to the cluster SED and thus be detectable.

We now consider the mass range of SMS that is expected in proto-GCs. Based on nucleosynthesis constraints, SMS with masses of $\sim 10^3 - 2\times 10^4$ \msun\  are needed (see Sect.\ \ref{s_sms_meth} and \citealt{pci17}).
The proto-GC+SMS spectral models presented above therefore correspond to the upper end, and the predicted magnitudes are accordingly probably at the bright end of our expectations. When we assume the same ratio $M_{\rm SMS}/M_{\rm cl}$ and a linear scaling of $L_{\rm SMS}$ with $M_{\rm SMS}$ from \cite{gieles18}, see also Fig.\ \ref{fig_ML}, this means that the proto-GCs with the minimum SMS mass would be approximately $\sim 20$ times (up to 3.25 mag) fainter than the predictions shown above. Presumably, we can expect the entire range of SMS/cluster masses, that is,\ also variations in the brightness of the proto-GCs in these early phases. The predicted colors are valid for any proto-GC/SMS combination that preserves the same $L_{\rm SMS}/L_{\rm cl}$ ratio as the cases shown here.

\begin{figure}[t]
\centering
\includegraphics[width=0.49\textwidth]{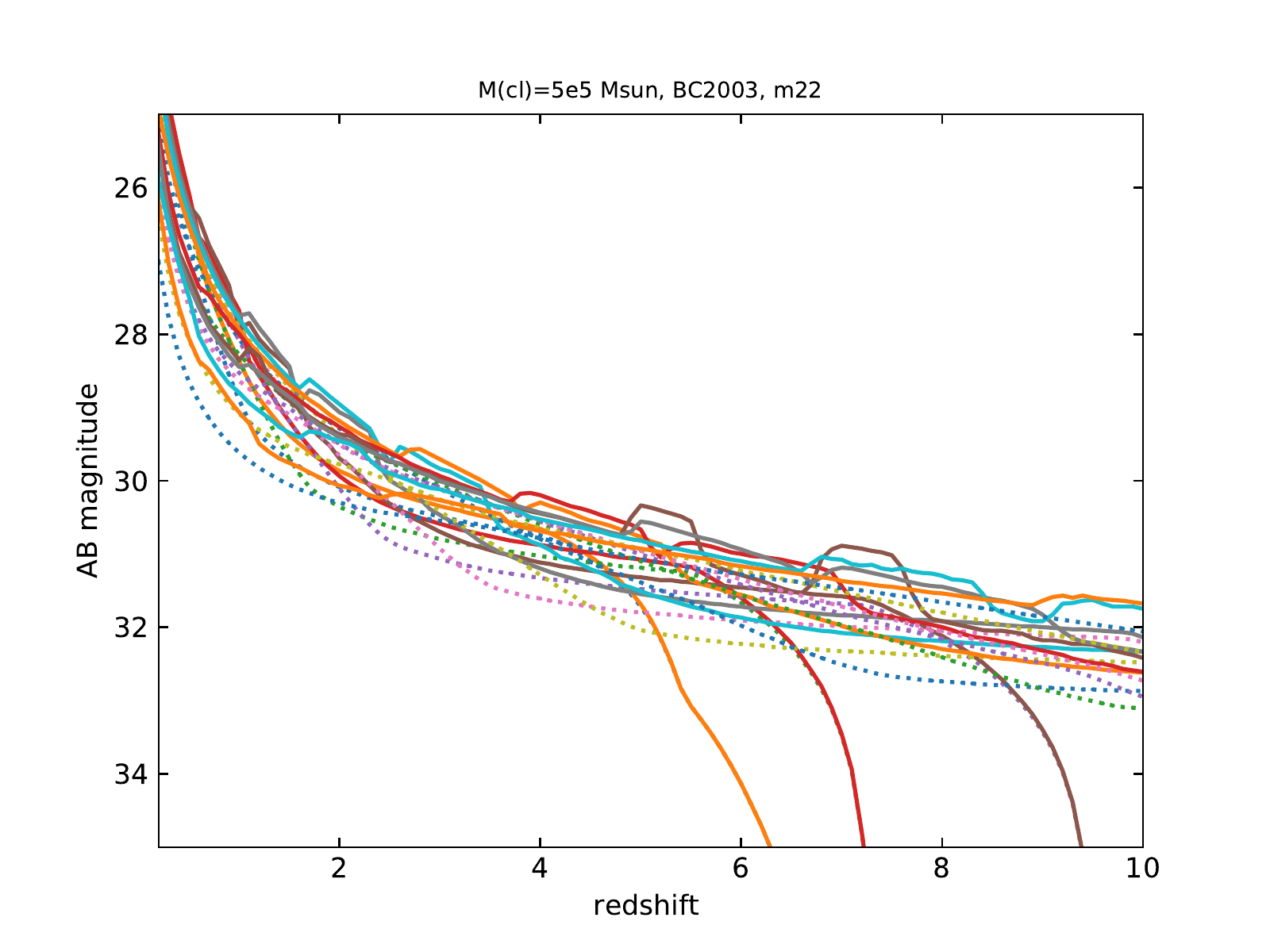}
\caption{Predicted magnitudes in the selected NIRCAM+MIRI filters
from 0.7-7.7 micron for young ($<4$ Myr) proto-GCs with $M_{\rm cl} = 5 \times 10^5$ \msun\ and negligible SMS contribution. Solid (dotted) lines show the predictions including (excluding) nebular emission. The three curves that drop out are due to the classical photometric drop-out from the shortest wavelength filters.}
\label{fig_mags_z_bcm22}
\end{figure}

If the SMS were quite compact, that is, if they\ had small radii corresponding to the case $\delta = 0.5$, for example, the integrated spectrum of proto-GCs would be indistinguishable from that of normal clusters of the same mass without SMS (cf.\ above). For example, for $M_{\rm SMS} \sim 10^3$ \msun\ the cluster mass is $M_{\rm cl} \sim 5 \times 10^5$ \msun\ 
\citep{gieles18}. In this case, the luminosity, magnitude, and colors of the proto-GC are the same as predicted above for the case of $\delta=1$ (cf.\ Table \ref{tab_clusters}), but excluding the SMS, that is,\ the pure cluster (or cluster+nebular) cases shown in Sect.\ \ref{s_starclu}. Typically, proto-GCs like this are expected to have magnitudes of $m_{\rm AB} \sim 31$ at $z \sim 6$, as illustrated in Fig.\ \ref{fig_mags_z_bcm22}.
Correspondingly, for an SMS that is ten times more massive, $M_{\rm cl}$ is approximately ten times higher, which then also holds for the proto-GC brightness.

\subsection{Comparison with other results}
\label{s_compother}

\cite{Renzini2017Finding-forming} and \cite{Pozzetti2019Search-instruct} have recently predicted photometry and number counts of GCs and discussed ways to find them in formation at high redshift. The underlying spectral models are from SSP models of  \cite{maraston05} and \cite{2003MNRAS.344.1000B}, which are the\ classical models used for cluster studies. \cite{Pozzetti2019Search-instruct} assumed a typical cluster mass of $2 \times 10^6$ \msun\ for their spectral predictions, which can be rescaled linearly to other cluster masses.
They neglected nebular emission, which significantly affects the SED for young ages ($< 10$ Myr) when the clusters are brightest in the UV (see, e.g.,\ Fig.\ \ref{fig_z7}).
As shown in Figure 12, this simple ISM effect can cause proto-GCs, even without SMS, to become brighter by $\sim 1$ mag in rest-UV and
more so at longer wavelengths. Numerous models that include nebular emission (continuum and also lines) exist in the literature \citep{sb99,sb99b,ygdrasil11,ygdrasil17}.
If a significant fraction of Lyman-continuum radiation were to escape the ISM of proto-GCs for some reason, the emission from the \hii\ region would be reduced correspondingly. Similarly, after the phase hosting massive ($\ga 10$ \msun) stars, nebular emission will be negligible and the SED predictions comparable to those of \cite{Pozzetti2019Search-instruct}.

\cite{Surace2018On-the-Detectio} and \cite{Surace2019On-the-detectio} have recently predicted observational properties of supermassive primordial stars, which could form under peculiar conditions and may be related to the formation of the first quasars.  The two main cases they examined correspond to accreting stars with total luminosities between $\log(L) \sim (4-5) \times 10^9$ \lsun\ and cool ($\sim 8000$ K) or hot (36000 K) temperatures, comparable to some of the SMS models examined here. Although a detailed comparison is clearly beyond the scope of this paper, we note that the predicted magnitudes
of our models A4 are comparable to those of the cool SMS of \citet{Surace2018On-the-Detectio}, which reach $\sim 28-31$ mag (AB) at $z \sim 6-10$ in photometric bands in common with our work. As expected, the more compact bluer primordial SMS are fainter in the same filters, and are accordingly more difficult to detect, according to \citet{Surace2019On-the-detectio}. Their spectral shape should also be less distinct from that of a normal stellar cluster, and they are therefore expected to be more difficult to identify if present, even in isolation, that is, without a surrounding young cluster population as considered here.

\subsection{Strategies for and feasibility of direct observations}

To identify the progenitors of present-day GCs in situ at high redshift, we should trivially identify compact clusters (sizes smaller than present-day sizes of a few tens of parsecs), which would in practice generally be unresolved, except maybe for strongly lensed sources. Beyond this simple criterion, there are no general features that would allow us to select proto-GCs in their initial phase, shortly after formation. The simplest approach assumes that the SED (and spectra) of proto-GCs can be described by that of normal SSPs, that is,\ ensembles of stars that formed at approximately the same time (instantaneously, or in a short burst), with a given metallicity, and a stellar mass typically above $\ga 10^{4-5}$ \msun, which is massive enough to form present-day GCs whose mass distribution peaks at $\sim 2 \times 10^5$ \msun\ \citep{harris2014}.

This method has been proposed by \cite{Renzini2017Finding-forming} and \cite{Pozzetti2019Search-instruct} to search for GCs in formation at high redshift. One of the caveats of this approach is that it does not take the possibly peculiar nature of proto-GCs in their initial or young phase into account, that is,\ multiple stellar populations and processes or sources that are responsible for these peculiarities.
The main justification for this inconsistent use of normal SSPs to describe the SED of young GCs is that we still do not fully understand their formation scenario. For a given scenario, such as the concurrent formation of an SMSs and GCs examined here, we have shown that the SEDs may show peculiar features that distinguish this scenario from that of normal SSPs.

With simple photometric redshift estimates (which can be obtained with the usual Lyman-break technique at high-$z$), the presence of an SMS in a proto-GC can concretely be recognised with two color measurements if the SMS is sufficiently bright and cool. In this case, the SED will resemble that of a composite stellar population, with normal stars surrounding the SMS that dominates the rest-UV flux, plus significant emission longward of the Balmer break, similar to that of a very bright cool star (see Figs. \ref{fig_A24} and \ref{fig_z7}). Some of the models examined here lead to a relatively blue UV spectrum (approximately constant in AB magnitudes, i.e.,\ a UV slope close to $\beta \sim -2$) plus a very strong Balmer break and strong emission in the rest-optical range (A2 in Fig.\ \ref{fig_A24}). Others can have a relatively red UV slope of $\beta \sim -0.8$ and be fainter again in AB magnitudes in the rest-optical range (A4). To distinguish these cases, a wavelength coverage that includes both the rest-UV and part of the optical is clearly needed. 

In short, although not all proto-GCs with SMS produce SEDs that significantly differ from SSPs, our results show that appropriately selected color measurements (requiring at least three photometric bands) are expected to allow us to observationally identify these objects.
In contrast, although spectroscopic observations could also help to confirm the presence of SMSs (cf.\ Sect.\ \ref{s_linesclu}), this does not appear feasible.

Imaging with the JWST will provide both a high spatial resolution to search for very compact or unresolved sources and a sufficiently wide spectral coverage to apply these search criteria. Furthermore, the proto-GCs with SMSs are predicted to be sufficiently bright (mag$_{\rm AB}$ $\sim$ 28-30  at $z \ga 6$ in the above cases) to be detectable with NIRCAM imaging. Observational searches for SMSs in young progenitors of GCs in the early Universe should therefore be possible, and if found, they would provide an important step toward our understanding of the formation of GCs and their multiple stellar populations.

The final state of SMSs in GCs remains a puzzle. If they explode as supernovae, it may be much easier to detect this explosion than the SMS itself. However, according to \citet{heger02}, it is likely that in the mass range of interest here, zero-metallicity SMSs experience pair instability and directly collapse into a stellar black hole (see also \citealt{2008A&A...477..223Y}, who studied the fate of solar metallicity stars up to $10^3$ \msun). We therefore do not expect any supernova in proto-GCs, at least at the metallicities considered here. Alternatively, SMS may disrupt under the effect of various types of instabilities (gravitational, pulsational, or general relativistic; \citealt{1959ApJ...129..637S,1964ApJ...140..417C,2008ApJ...684..212T,2008A&A...477..223Y,2013MNRAS.431.3036I}).   
No clear conclusion can currently be drawn either on the presence of intermediate-mass black holes in GCs, which is still a matter of debate \citep[e.g.,][]{2008ApJ...676.1008N,2017MNRAS.468.4429Z,2019MNRAS.482.4713Z,2018ApJ...862...16T,2018MNRAS.473.4832G,2019MNRAS.486.5008A}. Dynamical and evolutionary models for SMS that form through runaway collisions are clearly required to definitively predict their actual properties and their detectability at high redshift.

\section{Summary and conclusion}
\label{s_conc}
Motivated by the possible existence of SMSs that may have formed during the earliest phase of GC formation, as suggested by \citet{gieles18}, we have examined the observational properties of such stars with masses in the range of $\sim 1000-50000$ \msun\ and luminosities $\sim 10^{7.4} - 10^9$ \lsun\ and their detectability in the proto-GC phase.
To do so we computed computed non-LTE spherical  stellar atmosphere models that are appropriate for these extreme conditions with the \emph{CMFGEN} code \citep{hm98}, to predict the emergent spectra of supermassive low-metallicity stars. The models also cover a fairly wide range of radii, that is,\ effective temperatures from 7000 K to $\sim 130000$ K, which represent  SMS that are inflated by accretion during their growth phase or more compact very luminous stars. 

As expected for these high radiation field energy densities, we find strong non-LTE effects that significantly imprint the SED of these stars in the Lyman and the Balmer continuum. 
For example, we find
\begin{itemize}
\item a Balmer break in emission for cool SMSs with effective temperatures of about 10000 K. 
No such spectra are predicted for LTE models, and they are not known for normal stars. The high luminosities combined with the very low densities of the atmospheres imply strong non-LTE effects that also explain peculiar hydrogen line emission (Balmer lines in emission, higher series lines in absorption). At lower effective temperatures ($\sim$7000 K), the Balmer break shows a more classical absorption morphology. 
\item a Lyman break in emission for hot SMSs (\teff\ $>$ 40000 K), again due to strong non-LTE effects. For these stars, the classical infrared continuum excess observed in normal OB stars, and due to free-free emission in the extended atmosphere, is suppressed by the very low atmospheric density.
\end{itemize}

Using the predicted SMS spectra, we then examined the combined spectrum expected from young ($\la 3$ Myr) proto-GCs hosting such SMSs in the rest-UV, optical, and near-IR domain. For this we computed the expected SEDs of SMSs with different properties surrounded by a normal SSP that feeds the growth of the SMS, and whose properties are related to the SMS, as described in the scenario of \cite{gieles18}. We predict in particular the observed magnitudes of these sources over a wide redshift range ($z = 1 -10$) in the most important broadband filters of the JWST NIRCAM camera. The main results can be summarized as follows:
\begin{itemize}
    \item Inflated SMSs with M $\sim 10^4$ \msun, $\log L=9$ \lsun, and large radii, hence low effective temperatures, are predicted to outshine the proto-GC and have observed magnitudes \magab $\sim 28-30$ from $\sim 0.7-8$ \mum\ at $z \sim 4-10$, and brighter at lower redshift (see Fig.\ \ref{fig_mags_z}). 
    Proto-GCs with the same mass ($\sim 5 \times 10^5$ \msun) hosting less massive SMSs can be fainter by up to $\sim 2-3$ mag.
    Proto-GCs hosting SMSs should therefore be bright enough to be detected in very deep images, in particular with the upcoming JWST.
    \item The peculiar SED of cool (\teff $\la 10000$ K) and luminous SMSs implies that proto-GCs hosting such sources can be distinguished observationally from normal stellar clusters. This can be achieved with color-color diagrams probing the SED shortward and longward of the Balmer break (cf.\ Fig.\ \ref{fig_col_col}), for example. 
\end{itemize}

In short, our stellar atmosphere and spectral models show that SMSs that have been proposed to form during the short
initial (formation) phase of GCs can be bright enough and show peculiar SEDs to be detectable in situ{\em } at high redshift 
and be distinguishable from normal stellar clusters with very deep observations, such as imaging observations foreseen with the upcoming JWST. 
The formation scenario for GCs proposed by \cite{gieles18} should therefore be testable, and our predictions should provide a guide for searching 
for SMSs in such systems.

\section*{Acknowledgments}

We thank the referee, Nate Bastian, for a prompt and positive report. We warmly thank John Hillier for making the code \emph{CMFGEN} available to the community and for constant help with it. We thank Mark Gieles for useful comments on an early version of the paper. This work was supported by the "Programme National de Physique Stellaire" (PNPS) and the "Programme National Cosmologie et Galaxies" (PNCG) of CNRS/INSU co-funded by CEA and CNES.
CC acknowledges support from the Swiss National Science Foundation (SNF) for the project 200020-169125 "Globular cluster archeology".
LH was sponsored by the Swiss National Science Foundation (project number 200020-172505).

\bibliographystyle{aa}
\bibliography{sms_gc}

\begin{appendix}

\section{Effect of surface gravity on  the Balmer jump}
\label{ap_Balmer_logg}

\begin{figure}[]
\centering
\includegraphics[width=0.49\textwidth]{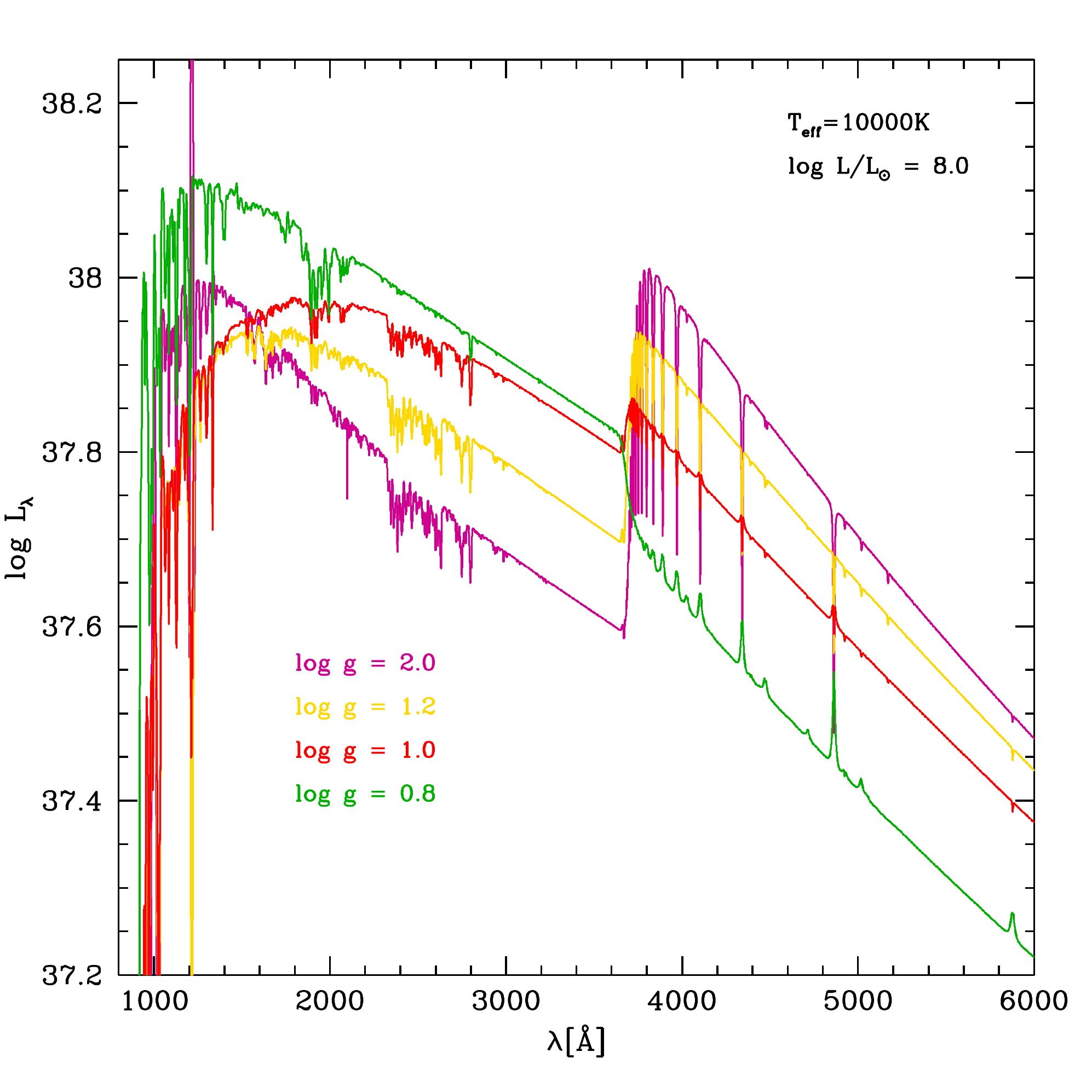}\\
\caption{SED of models with \teff\ = 10000 K, \lL\ = 8.0 and different surface gravities.}
\label{fig_t10L8}
\end{figure}

In Fig.\ \ref{fig_t10L8} we show the effect of the surface gravity on the SED of models with \teff\ = 10000 K  and \lL\ = 8.0. When \logg\ decreases, the Balmer break size is reduced, and the break switches from absorption to emission only for the lowest \logg\ model. This is understood by the stronger non-LTE effects in low surface gravity (and thus low-density) models; see Sect.\ \ref{phy_bal}.  

\section{Hydrogen lines at \lL\ = 9.0}
\label{ap_HlinesL9}

\begin{figure*}[hbt]
\centering
\includegraphics[width=0.49\textwidth]{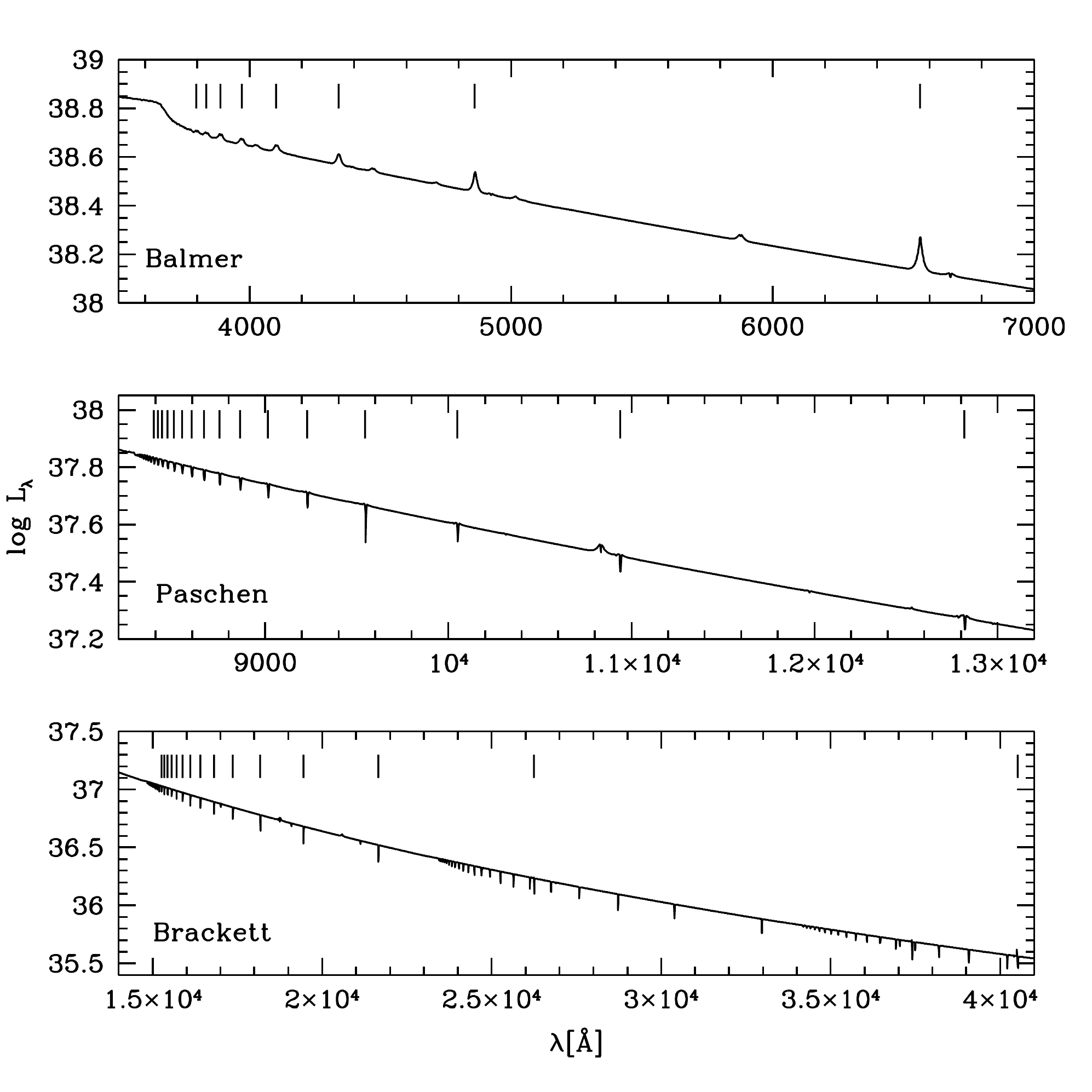}
\includegraphics[width=0.49\textwidth]{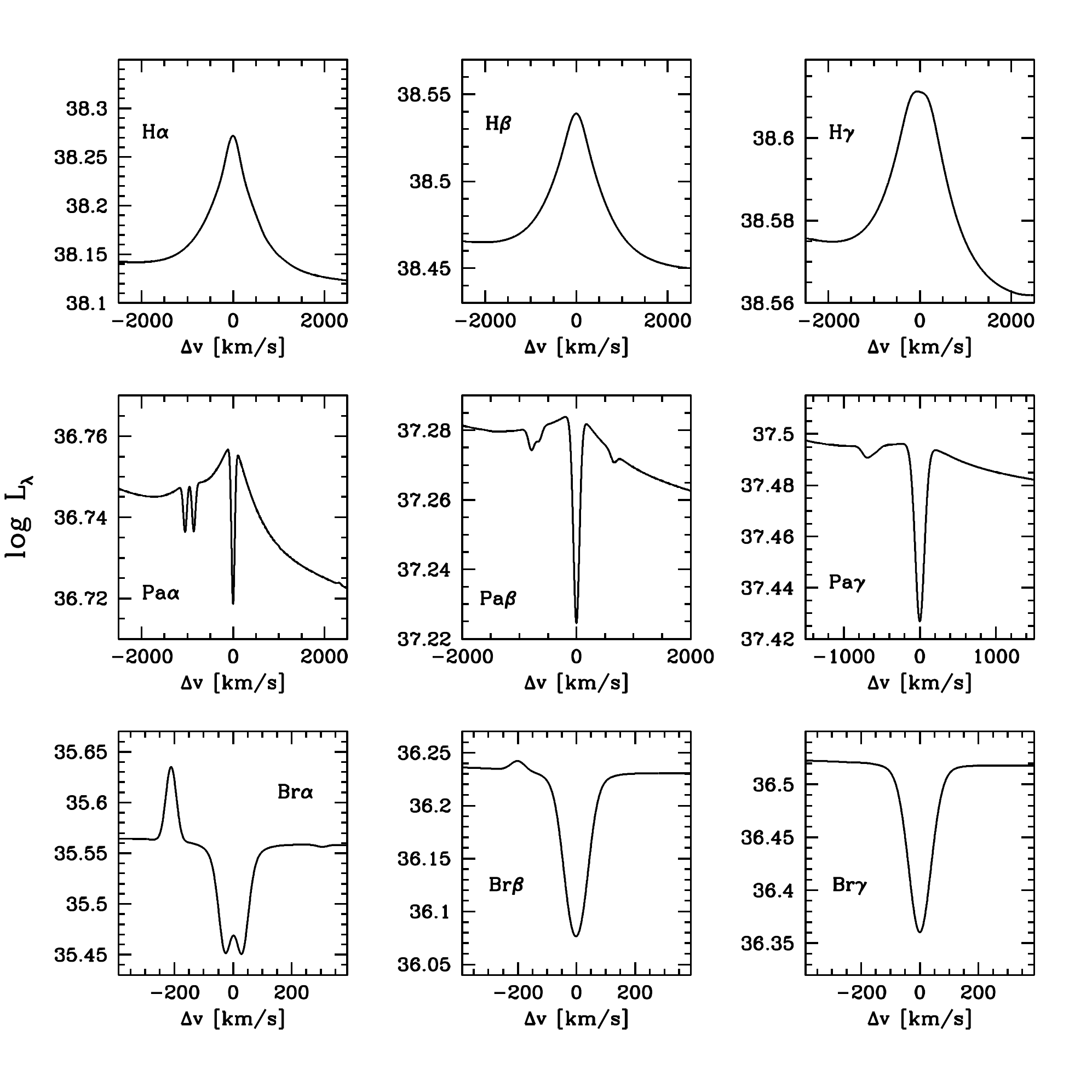}\\
\caption{Same as Fig.\ \ref{fig_Hseries} for the model with \lL\ = 9.0.}
\label{fig_HseriesL9}
\end{figure*}

Fig.\ \ref{fig_HseriesL9} shows the Balmer continua and the first lines of the Balmer, Paschen, and Brackett series for the model with \teff\ = 10000 K, \logg\ = 0.8, and \lL\ = 9.0 (model A4). 

\section{Physics of models B2 and C2}
\label{ap_hot}

\begin{figure*}[hbt]
\centering
\includegraphics[width=0.32\textwidth]{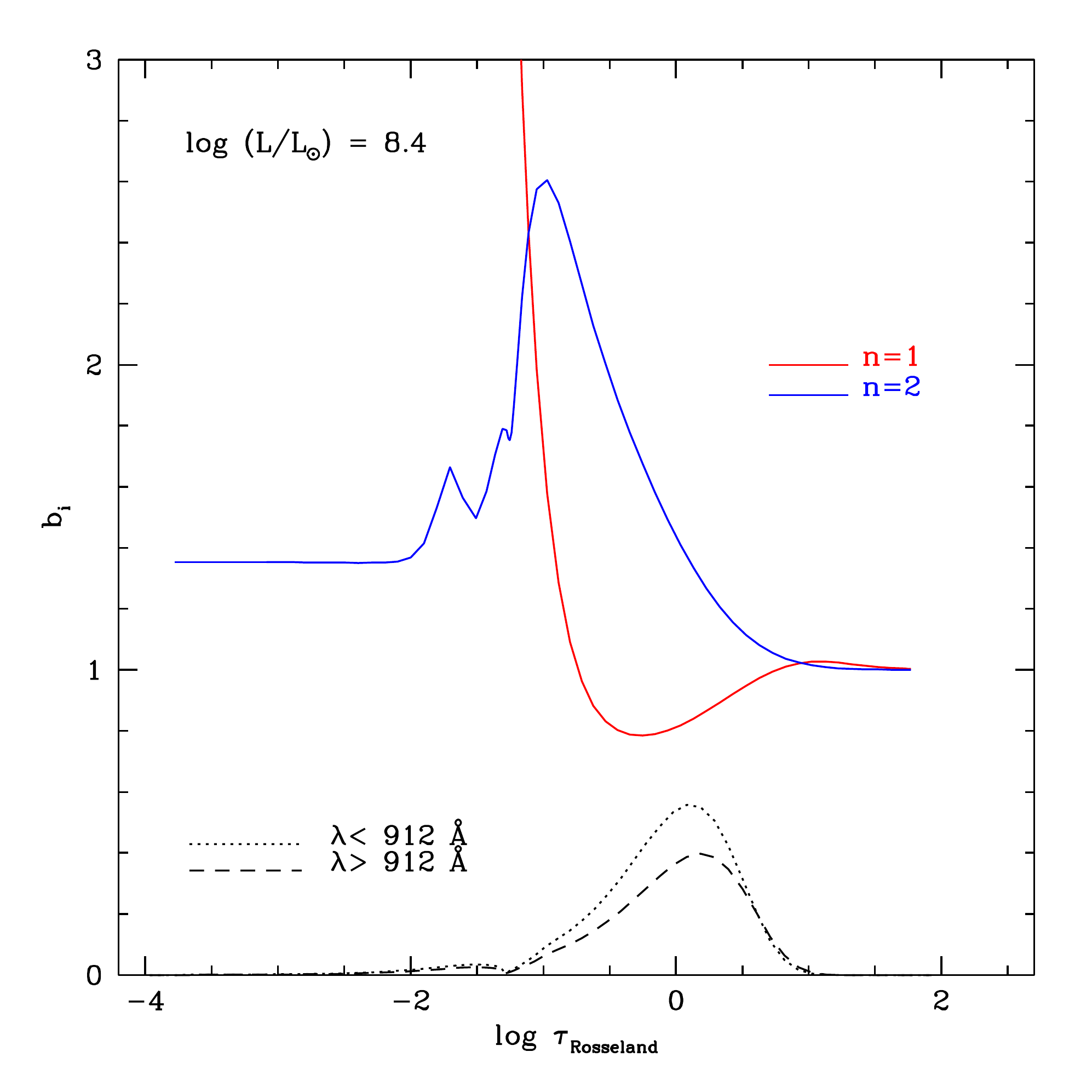}
\includegraphics[width=0.32\textwidth]{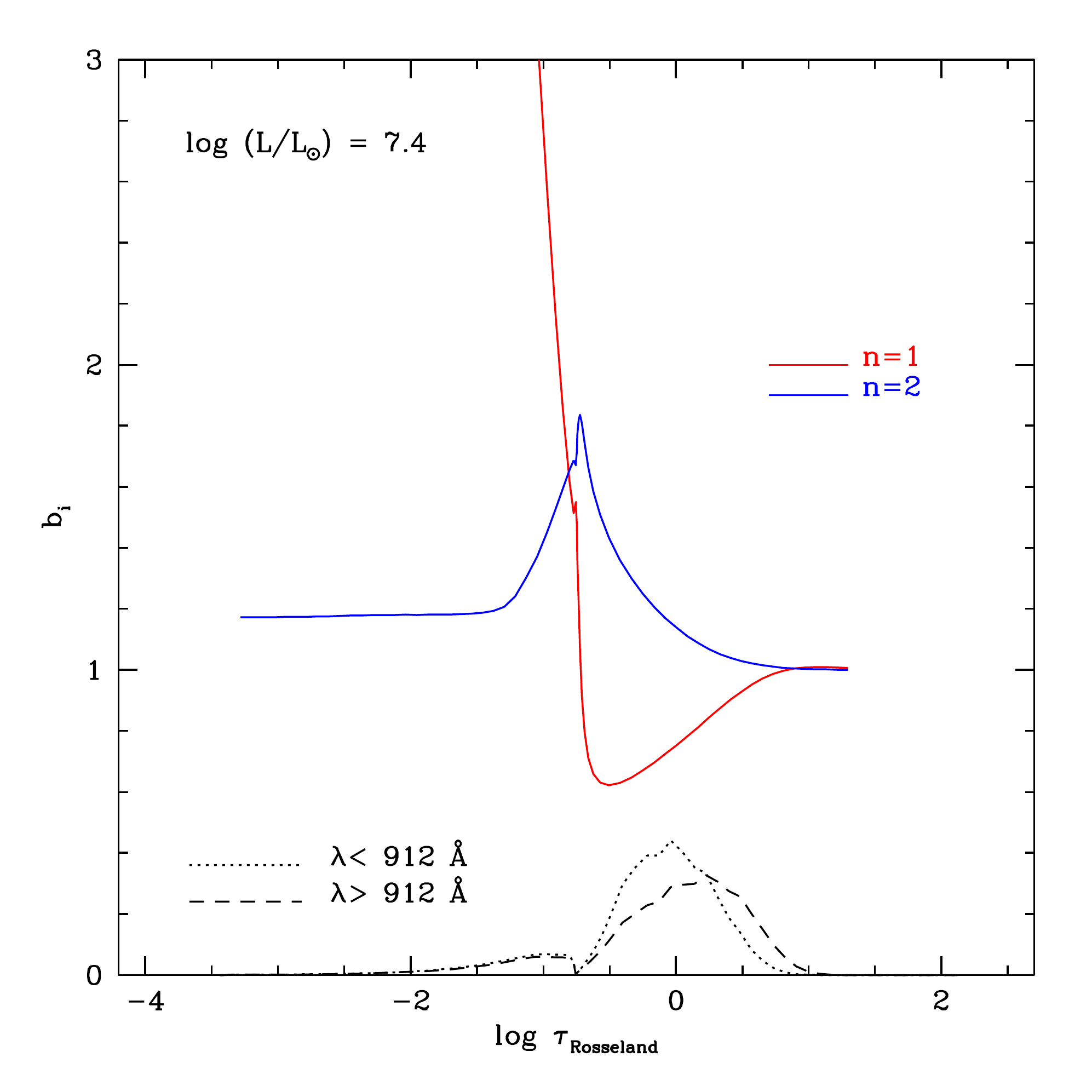}
\includegraphics[width=0.32\textwidth]{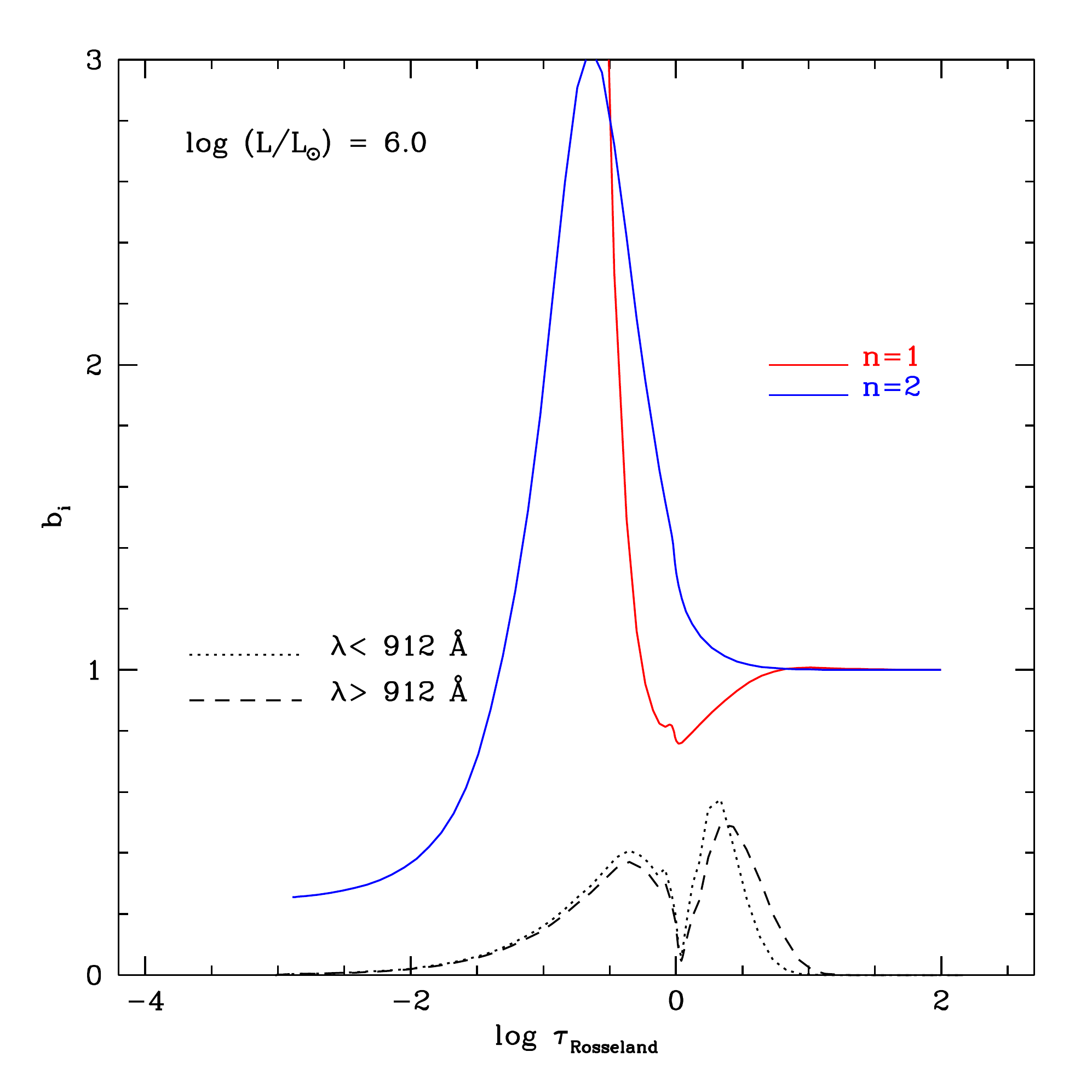}
\caption{Departure coefficients for the ground and first excited levels of hydrogen for model B2 (left), C2 (middle), and a model with \lL\ = 6.0 (right).}
\label{fig_bit43}
\end{figure*}

Fig.\ \ref{fig_bit43} shows the departure coefficients of the models shown in Fig.\ \ref{fig_t43}. Over the continuum-formation region, the ground level of the model with \lL\ = 6.0 (representing a normal massive star, and referred to as the reference model below) is both over- and underpopulated, depending on the optical depth. In the more luminous models, the ground level is underpopulated over the formation region. In all models, the first excited level (n=2) is always overpopulated. 

\begin{figure}[hbt]
\centering
\includegraphics[width=0.49\textwidth]{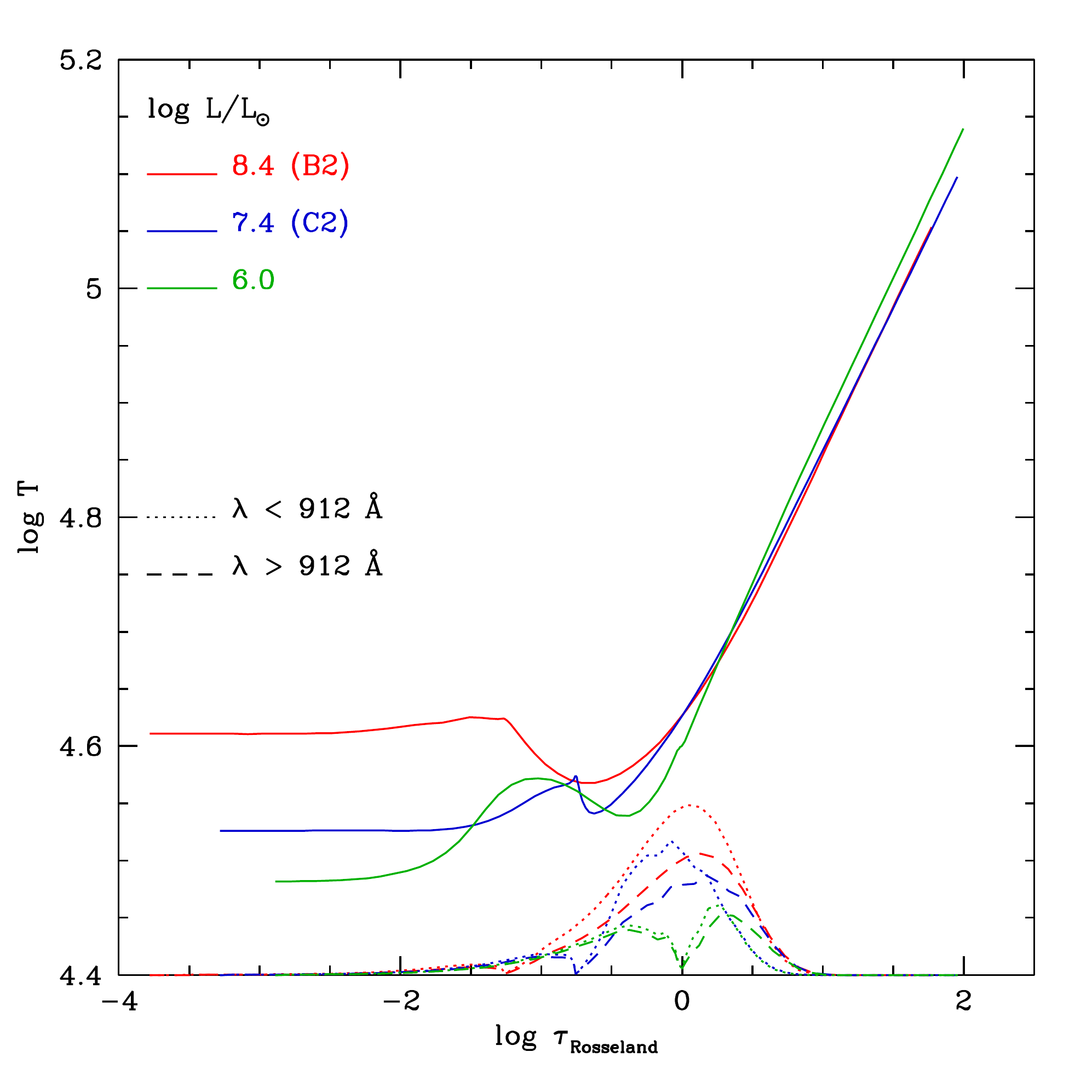}
\caption{Temperature structures of models B2, C2, and a model with the same parameters but \lL\ = 6.0. The formation regions of the continuum below and above the Lyman break are shown by dashed and dotted lines.}
\label{fig_struT43}
\end{figure}

Fig.\ \ref{fig_struT43} shows the temperature structures of the same models. In the reference model and in model B2, the continua below and above the Lyman break are formed in the same region. Consequently, the difference in the shape of the break is mainly dominated by the effects of departure coefficients: an average lower population of the ground level favors a stronger emission on the blue side of the break. In model C2, the continuum shortward of the break is formed above the continuum shortward of the break (i.e., formed at a higher height in the atmosphere). The temperature effect (Eq.\ \ref{eq_sb}) dominates and the jump is smaller than in model B2. 
 
\begin{figure}[hbt]
\centering
\includegraphics[width=0.49\textwidth]{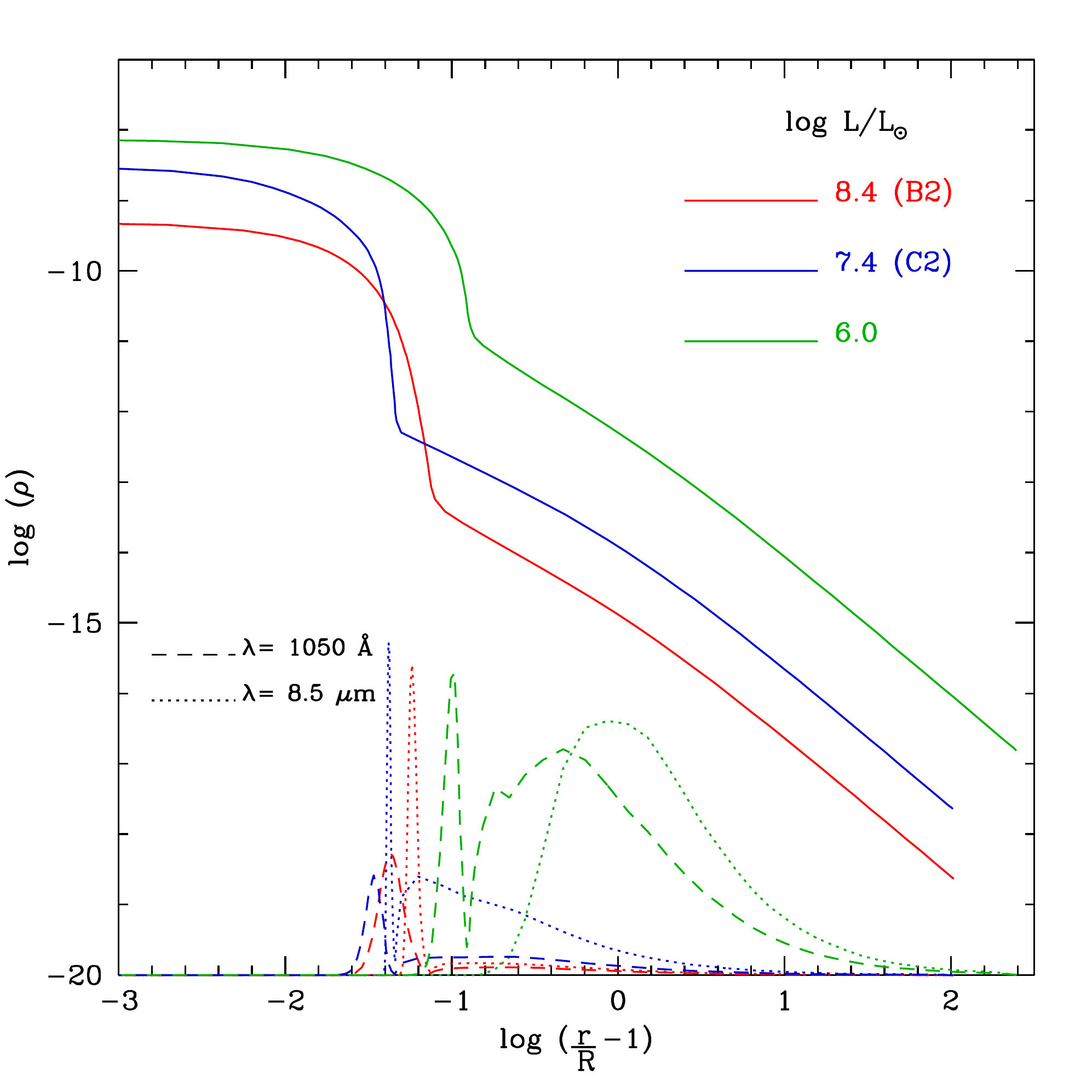}
\caption{Density structures of models B2, C2, and a model with the same parameters but \lL\ = 6.0 as a function of height in the atmosphere (R is the stellar radius). The formation regions of the continuum at 1050 \AA\ and 8.5 \mum\ are shown by dashed and dotted lines.}
\label{fig_strurho43}
\end{figure}

Infrared excess in the spectra of massive stars is due to free-free emission in the ionized wind \citep{dk72,wb75}. The total SED can be viewed as the sum of the stellar SED and an emission component formed in the extended atmosphere that resembles an HII region.
The explanation of the absence of emission excess at long wavelength in the high-luminosity models is given by Fig.\ \ref{fig_strurho43}. In the reference model, the stellar continuum (taken at 1050 \AA\ for illustration) is formed over a wide region, but with a significant contribution from the photosphere (i.e., below the sudden drop in density). The free-free emission from the extended wind (evaluated at 8.5 \mum) is emitted at heights that are about an order of magnitude higher. Consequently, the volume corresponding to this emission region is large and the free-free emission dominates the stellar emission. In model B2, the free-free emission from the wind and the stellar flux are emitted in the same region. The former is negligible compared to the latter because the size of the wind-emitting region is small. The main reason for this different behavior is the low density in the high-luminosity atmosphere (see the densities in Fig.\ \ref{fig_strurho43}): most of the wind is optically thin. Only close to the photosphere does the density reach a level that is sufficient to produce emission (the infrared continuum is emitted in regions where $\log \rho \sim\ -13.0$ in all models). Model C2 is somewhat intermediate, with a small amount of free-free emission.

\section{Synthetic photometry}
\label{ap_photom}

The synthetic photometry computed for some of the main filters of the NIRCam and MIRI instruments on the JWST for this study and presented in the figures of Sect.\ \ref{s_photom} and \ref{s_discuss} is made available in electronic format at the CDS.
Table \ref{tab_photom} lists the available combinations of objects (SMS, cluster, and SMS+cluster) and indications on the absence or inclusion of nebular emission for which integrated photometry is published. Each of Tables 4 to 11 contains the following information: Column 1 lists the redshift, Column 2 gives the AB magnitude in the F070W filter, and all subsequent columns list the same for the following filters: F090W, F115W, F150W, F200W, F277W, F356W, F410W, F444W, F560W, and F770W. In total, AB magnitudes are listed for 11 broadband filters.

\begin{table}
  \caption{Synthetic photometry for different SMS models, cluster, and the total flux, presented in this work, computed in the NIRCAM JWST filters. The entry gives the table number of the tables, which are available in electronic format at the CDS.} 
  \label{tab_photom}
\begin{tabular}{lccccc}
\hline
        object  & object  &  object & total (SMS+cluster) \\
        & (no nebular) &  (with nebular) & (with nebular) \\
\hline
A2 &       D.1  & - & D.2 \\
A4 &       D.3  & - & D.4 \\
B3 &       D.5  & - & D.6 \\
cluster &  D.7 & D.8 & - \\
 \hline                                                                   
\end{tabular}                                                            
\end{table}

\end{appendix}


\end{document}